\documentclass[twocolumn,apj,iop,twocolappendix]{emulateapj}

\usepackage{amssymb,amsmath}
\usepackage{mathrsfs}
\usepackage{times}
\usepackage{color}

\newcommand{\teff}{\ensuremath{T_{\rm eff}}}
\newcommand{\beq}{\begin{equation}}
\newcommand{\eeq}{\end{equation}}
\newcommand{\intd}{{\rm d}}
\newcommand{\msun}{\ensuremath{M_\odot}}
\newcommand{\tp}{\ensuremath{t_{\rm P}}}
\newcommand{\tw}{\ensuremath{t_{\rm w}}}
\newcommand{\texp}{\ensuremath{t_{0}}}
\newcommand{\ergs}{ergs\,s$^{-1}$}
\newcommand{\bol}{\ensuremath{_{\rm bol}}}
\newcommand{\lbol}{\ensuremath{L\bol}}
\newcommand{\fbol}{\ensuremath{F\bol}}
\newcommand{\kms}{km\,s\ensuremath{^{-1}}}
\newcommand{\mni}{\ensuremath{M_{\rm Ni}}}
\newcommand{\tc}{\ensuremath{T_{\rm c}}}
\newcommand{\rv}{\ensuremath{\mathscr{R}_V}}
\newcommand{\ri}{\ensuremath{\mathscr{R}_i}}
\newcommand{\swift}{{\em Swift}}
\newcommand{\ndof}{\ensuremath{N_{\rm DOF}}}

\begin{document}
\title{A Global Model of The Light Curves and Expansion Velocities of Type II-Plateau Supernovae}
\shorttitle{A Global Model of Type II-Plateau Supernovae}
\shortauthors{Pejcha \& Prieto}

\author{ Ond\v{r}ej Pejcha\altaffilmark{1}}
\affil{Department of Astrophysical Sciences, Princeton University, 4 Ivy Lane, Princeton, NJ 08540, USA}
\email{pejcha@astro.princeton.edu}
\and
\author{Jose L. Prieto}
\affil{N\'ucleo de Astronom\'ia de la Facultad de Ingenier\'ia, Universidad Diego Portales, Av. Ej\'ercito 441, Santiago, Chile}
\affil{Millennium Institute of Astrophysics, Santiago, Chile}
\altaffiltext{1}{Hubble and Lyman Spitzer Jr.\ Fellow}

\begin{abstract}
We present a new self-consistent and versatile method that derives photospheric radius and temperature variations of Type II-Plateau supernovae based on their expansion velocities and photometric measurements. We apply the method to a sample of $26$ well-observed, nearby supernovae with published light curves and velocities. We simultaneously fit $\sim 230$ velocity and $\sim 6800$ magnitude measurements distributed over $21$ photometric passbands spanning wavelengths from $0.19$ to $2.2\,\mu$m. The light curve differences among the Type II-Plateau supernovae are well-modeled by assuming different rates of photospheric radius expansion, which we explain as different density profiles of the ejecta and we argue that steeper density profiles result in flatter plateaus, if everything else remains unchanged. The steep luminosity decline of Type II-Linear supernovae is due to fast evolution of the photospheric temperature, which we verify with a successful fit of SN1980K. Eliminating the need for theoretical supernova atmosphere models, we obtain self-consistent relative distances, reddenings, and nickel masses fully accounting for all internal model uncertainties and covariances. We use our global fit to estimate the time evolution of any missing band tailored specifically for each supernova and we construct spectral energy distributions and bolometric light curves. We produce bolometric corrections for all filter combinations in our sample. We compare our model to the theoretical dilution factors and find good agreement for the $B$ and $V$ filters. Our results differ from the theory when the $I$, $J$, $H$, or $K$ bands are included. We investigate the reddening law towards our supernovae and find reasonable agreement with standard $\rv \sim 3.1$ reddening law in $UBVRI$ bands. Results for other bands are inconclusive. We make our fitting code publicly available.
\end{abstract}

\section{Introduction}

Core-collapse supernovae announce the death of (at least some) stars with initial masses $\gtrsim 8\,\msun$. The gravitational collapse of the Chandrasekhar-mass iron or oxygen-neon-magnesium core of these massive stars rebounds when the strong nuclear interaction causes a stiffening of the equation of state. A shock wave then propagates outwards into the infalling matter, but simulations show that it halts its progress and turns into a quasi-static accretion shock. With the aid of neutrinos emanating from the nascent proto-neutron star and some additional poorly understood mechanism, the shock starts moving out again and produces a supernova explosion \citep[e.g.][]{janka12,ugliano12,burrows13}. Although exciting, the prospects of directly constraining the supernova explosion mechanism through observations of neutrino emission, gravitational waves or their combination \citep[e.g.][]{ott04,ott12,yuksel07,muller14} are extremely uncertain, because the supernova must explode nearby. Instead, constraints on the explosion mechanism can be obtained by trying to understand observed patterns in the explosions, their remnants, and their progenitors.

As the reinvigorated shock propagates through the progenitor, it heats up the swept up material. When the temperatures are higher than about $5\times 10^9$\,K, heavy elements such as $^{56}$Ni are produced \citep[e.g.][]{weaver80,woosley88,thielemann90}. For lower temperatures, only lighter elements are synthesized and still lower temperatures cause only ionization of the matter. After the short initial luminosity spike when the supernova shock breaks out of the surface of the star \citep[e.g.][]{ensman92,chevalier08,soderberg08,katz10,nakar10,tominaga11}, the luminosity increases as the surface area of the ejecta expands. Simultaneously, the effective temperature decreases. If the progenitor star had a substantial hydrogen envelope, the ejecta are optically-thick and the combination of homologous expansion and the decreasing effective temperature results in a phase of relatively constant optical brightness; we observe a Type II-P supernova. When the ejecta becomes optically-thin, the brightness drops and the luminosity evolution becomes dominated by the deposition of energy of the decaying radioactive nuclei. Specifically, the normalization of the exponentialy decreasing luminosity is related to the total mass of $^{56}$Ni that was synthesized during the explosion. 

Theoretically, the supernova explosion energy combines the contributions from the binding energy of the progenitor, the neutrino-driven wind, recombination of the dissociated nuclei, and exothermic nuclear burning, all of which happen predominantly deep in the progenitor close to the proto-neutron star \citep[e.g.][]{scheck06,ugliano12}. Supernova explosion energies are estimated by comparing the observed fluxes and spectra to theoretical models of expanding supernova atmospheres with assumptions on the progenitor properties and the explosion remnant \citep[e.g.][]{arnett80,litvinova83,litvinova85,hamuy03,utrobin08,kasen09}. In recent years, a number of red supergiant progenitors to Type II-P supernovae have been identified and their initial masses have been estimated \citep[e.g.][]{vandyk03,vandyk12a,vandyk12b,smartt04,smarttetal09,li06,mattila08,fraser12,maund14}. One surprising aspect of these discoveries is that there appears to be a lack of high-mass red supergiants exploding as Type II-P supernovae \citep{li06,kochanek08,smartt09,smarttetal09}. Although there are several explanations \citep{yoon10,walmswell12,kochanek12,groh13}, an exciting possibility is that some massive stars do not explode as ordinary supernovae but instead collapse to a black hole, potentially accompanied only by a weak transient \citep{nadezhin80,kochanek08,lovegrove13,piro13,kochanek14a,kochanek14b,horiuchi14}. The supernovae with estimates of progenitor mass, explosion energy and nickel mass can potentially offer strong constraints on the supernova explosion mechanism.

The Type II-P supernovae have also been used to measure distances using either the expanding photosphere method \citep[e.g.][]{kirshner74,schmidt94a,hamuy01,baron04,dessart05,dessart06,jones09} or the ``standardized candle'' method  \citep[e.g.][]{hamuy02,nugent06,poznanski09,poznanski10,dandrea10,olivares10}. The expanding photosphere method uses theory-based ``dilution'' factors to transform the black-body flux of an observed color temperature derived from a given filter set to the total supernova flux. The theory necessary to calculate the dilution factors includes line blanketing and non-LTE effects in an expanding medium. The distance is estimated by comparing the apparent angular radius of the supernova with expansion velocities under the assumption of homologous expansion. The standardized candle method employs a correlation between the supernova luminosity and the measured expansion velocity to infer the distances. While they are still not competitive with Type Ia supernovae for cosmology, Type II-P SNe offer an independent and promising method for estimating distances with different astrophysics and systematics.

In practice, the extraction of supernova distances, explosion energies, nickel yields and progenitor masses from observations is not straightforward. Most of all, the light from the supernova or the progenitor is extinguished by gas and dust around the star, in its host galaxy and in our Galaxy. The extinction can be estimated from the maps of galactic dust \citep{schlegel98}, the spectral lines of the intervening gas imprinted on the supernova spectrum \citep[e.g.][]{munari97,poznanski11,poznanski12}, by comparing the supernova colors to well-observed template supernovae \citep{olivares10}, or by modelling of the supernova spectra in the well-defined continuum windows \citep{dessart06,dessart08,baron07}. All of these recipes provide potentially biased results for the explosion progenitor masses, explosion energies and nickel mass from the heterogeneous data of different supernovae. Furthermore, \citet{faran14} tested several commonly used methods for dust-extinction correction and argued that none of them increase the uniformity of the sample.

To expand on our previous effort to understand the supernova progenitors, the explosion mechanism and its observational signatures \citep{prieto08a,prieto08b,prieto08c,prieto12,prieto13,pt12,pt14,pejcha_ns,pejcha_cno} and to address the uncertainties and inhomogeneities in determining Type II-P supernova parameters, we present a new method to fit the multi-band light curves and expansion velocities of Type II-P supernovae. Our method combines the founding principles of expanding photosphere method with the generalization of the Baade-Wesselink approach in Cepheids \citep{pejcha12} to decompose the observed multi-band light curves and expansion velocities of Type II-P supernovae into radius and temperature variations. The changes in the photospheric radius are constrained by the expansion velocities under the assumption of homologous expansion and affect all observed photometric bands in the same achromatic manner. We assume that the changes in the spectral energy distribution of supernovae are driven by variations in a single underlying parameter, the temperature. This represents the chromatic part of the light curve, which enters differently for each photometric band. The parameters necessary to project the observed light curves and velocities on radius and temperature changes are obtained in the fitting process. Our method does not require any input from theoretical supernova atmosphere models. In fact, and importantly, our results can constrain these models. Furthermore, we fit each photometric band independently, which assures that supernova observed in any combination of our photometric bands can be successfully fitted and its parameters estimated.

In this paper, we present the model and the first findings based on the limited set of publicly-available data. In Section~\ref{sec:model}, we present the basic equations of the model, identify the degeneracies and design priors to ensure reasonable fits, describe the available data and the fitting methods, and discuss the relation to previous methods. In Section~\ref{sec:results}, we present the fits to the light curves along with the range of applicability of the model, discuss the morphology of the light curves, and present the distance estimates to individual supernovae. We also construct spectral energy distributions, bolometric light curves and corrections, and estimate nickel masses of individual supernovae. We compare our results to the theoretical supernova atmosphere models by calculating the dilution factors. In Section~\ref{sec:disc}, we discuss the reddening law and constraints from our model. We also describe possible extensions to our model. In Section~\ref{sec:conclusions}, we summarize our findings. In the Appendix, we describe our implementation of priors on the model.

\section{Model}
\label{sec:model}


We model the supernova magnitude $m_i(t)$ in a photometric band $i$ at a time $t$
\beq
m_i(t) = \overline{M}_i +\mu_j + \mathscr{R}_iE(B-V) - 2.5\Pi(t) - 2.5\Theta_i(t),
\label{eq:main}
\eeq
where $\overline{M}_i$ is the absolute magnitude in band $i$ of the object at a distance of $10$\,pc, $\mu_j$ is the distance modulus of the host galaxy $j$, $\ri$ is the ratio of total to selective extinction, $E(B-V)$ is the total reddening, $\Pi$ represents changes in the luminosity due to changes in the surface area at constant temperature and other achromatic luminosity changes, and $\Theta_i$ describes the chromatic magnitude changes in band $i$ due to temperature changes. In this work, we assign a single value of $E(B-V)$ to each supernova and assume that there is a universal extinction law $\ri$, which we discuss in Section~\ref{sec:reddening}.  We also emphasize that our $E(B-V)$ is the total reddening that includes the Galaxy and the supernova host galaxy contributions. We do not pre-correct the observations for Galaxy reddening, because this could potentially introduce additional unknown systematical errors in our analysis and we are interested in the total photometric reddening for our supernovae.

To apply Equation~(\ref{eq:main}) in the supernova context, we assume that after the moment of explosion $\texp$, the optical emission comes predominantly from a photosphere in the optically-thick expanding medium with a relatively well-defined radius $R$ and temperature $T$, the ``plateau''. We assume that the expanding medium is homologous so that $R$ is related to the spectroscopically measured expansion velocity $v$ as
\beq
R = v(t-\texp),
\label{eq:homo}
\eeq
where we are neglecting the initial radius of the progenitor star. Here, $\texp$ is the zero-point time for each supernova, which might be systematically offset from the true moment of explosion due to missing physics in our model. We assume that the line used to infer expansion velocity is directly related to the photosphere so that Equation~(\ref{eq:homo}) is valid. Expansion velocities are modeled as
\beq
v = \omega_0 (t-\texp)^{\omega_1}+\omega_2,
\label{eq:v}
\eeq
where $\omega_0$, $\omega_1$, $\omega_2$ are parameters. For $v$ in \kms and $t$ in days, the reference magnitude $\overline{M}_i$ corresponds to an object with $R_0=8.64\times 10^{9}$\,cm ($\Pi=0$). We parameterize the change in the spectral energy distribution as the supernova evolves using a parameter $\tau$. Motivated by observational studies of supernovae during this phase, we model the time evolution of $\tau$ as a linear decay
\beq
\tau_{\rm thick} = \alpha_0 (t-\texp) + \alpha_1.
\label{eq:tauthick}
\eeq
Over time $\tw$, the plateau smoothly transitions to an optically-thin phase, where the optical luminosity comes from the deposition of energy from the decay of radioactive nuclei. Motivated by observations \citep[e.g.][]{hamuy01}, we model this part of the evolution as an achromatic exponential luminosity decay with $\tau \equiv \tau_{\rm thin}$ constant in time. Improvements in the treatment of the exponential decay will be a subject of future work, but in this paper, we focus predominantly on the optically-thick plateau phase.

We combine the photospheric phase and the exponential decay phase as
\begin{eqnarray}
\Pi &=& \log_{10}\left\{(1-w)R^2  + w10^{\gamma_0 (t-\texp -\tp)+\gamma_1}   \right\},\\
\tau &=& (1-w)\tau_{\rm thick} +  w\tau_{\rm thin},
\end{eqnarray}
where $\gamma_0$ and $\gamma_1$ parameterize the exponential decay, $\tp$ is the duration of the photospheric phase, and $w$ is the weight function
\beq
w = \left[1+ \exp \left(\frac{t-\texp-\tp}{\tw} \right)  \right]^{-1},
\label{eq:w}
\eeq
which we choose after \citet{olivares10} to resemble the typical transition phase in Type II-P supernovae. Equation~(\ref{eq:w}) smoothly connects the photospheric and exponential decay phases, which allows for the convergence of the fitting routine. Equation~(\ref{eq:w}) implies that $\tp$ is defined at the midpoint between the end of the plateau and the beginning of the exponential luminosity decay, similarly to \citet{hamuy03}. This is somewhat different from other recent studies that measure the duration of the photospheric phase at the end of the plateau \citep{anderson14a,sanders14}.

To calculate the magnitude at any given time, we need to know how temperature changes affect individual photometric bands through $\Theta_i$. We model the temperature dependence of $\Theta_i$ as a low-order polynomial in $\tau$
\beq
\Theta_i = \sum_{n=1}^3\frac{1}{n!}\beta_{n,i}\tau^n
\label{eq:theta}
\eeq
with a matrix of coefficients $\beta_{n,i}$. In principle, $\Theta_i$ depends also on metallicity and other parameters, however, since the $\beta_{n,i}$ will be obtained by fitting the data, these additional dependencies will matter only when comparing fits of individual objects. Metallicity variations between supernovae will be projected on other parameters such a $\alpha_0$, $\alpha_1$, and $E(B-V)$, making the metallicity signature much smaller than one would expect \citep{pejcha12}. In our previous work on Cepheids \citep{pejcha12}, we found that the metallicity term is typically few percent of the $\beta_{1,i}$. We believe that this is also the case for supernovae. We will apply our model to local low-redshift supernovae with good observational coverage. As a result, we do not include K-corrections. Changes to our model in this direction would be an obvious extension.

Finally, we want to address the meaning of the parameter $\tau$ that we have so far left unexplained. The spectral energy distribution changes primarily due to changes in the photospheric temperature, however, in the model presented here, $\tau$ is neither the effective nor the color temperature of the supernova. Instead, it is advantageous to think about $\tau$ as a parameter that describes changes in the spectral energy distribution and that its meaning is limited by the model presented above. We show later in Section~\ref{sec:bc} that $\tau$ is approximately linearly proportional to the logarithm of the effective temperature. Alternatively, we could have forced $\tau$ to be identical to the effective temperature by integrating the spectral energy distribution. However, this approach would mean that the definition of $\tau$ changes when a new band is added.

\subsection{Degeneracies and priors}
\label{sec:priors}

The parameters of the model are obtained by minimizing the master constraint 
\beq
\mathcal{H} = \sum_{\rm all\ data}\left(\frac{m^{\rm obs}-m}{\sigma}\right)^2 +\sum_{\rm all\ data}\left(\frac{v^{\rm obs}-v}{(1-w)^{-1}\sigma}\right)^2 + S,
\label{eq:constraint}
\eeq
where $m^{\rm obs}$ and $v^{\rm obs}$ are observed magnitudes and expansion velocities with uncertainties $\sigma$, $S$ includes contributions from all priors, and the sums are over all supernovae, photometric bands, and measurements. We modify the uncertainties of the expansion velocities by a factor of $1-w$ to lower the fitting weight of velocity measurements during the time when the supernova ejecta is becoming transparent. We collectively denote the first two terms in Equation~(\ref{eq:constraint}) as $\chi^2$.

The model in Equation~(\ref{eq:constraint}) is hierarchical in the sense that it includes parameters pertaining only to a single supernova, ($\texp$, $\tp$, $\tw$, $E(B-V)$, $\omega_0$, $\omega_1$, $\omega_2$, $\alpha_0$, $\alpha_1$, $\gamma_0$, $\gamma_1$, $\tau_{\rm thin}$), only to the host galaxies ($\mu_j$), and globally to all data ($\overline{M}_i$, $\ri$, $\beta_{n,i}$). Although we fit Equation~(\ref{eq:constraint}) to a large amount of data, some of the model parameters cannot be constrained completely independently. We address these issues by fixing some parameters and adding priors. Equation~(\ref{eq:main}) implies that the model magnitude $m_i$ remains unchanged if we move all objects by $\mu \rightarrow \mu+\Delta\mu$ while simultaneously changing $\overline{M}_i\rightarrow \overline{M}_i-\Delta\mu$. We fix this degeneracy by fixing the distance modulus to M95, the host galaxy of SN2012aw, to $\mu_{\rm M95} \equiv 30.00$\,mag. This number is very close to the weighted average of $30.01$\,mag from the NASA Extragalactic Database (NED), which includes numerous distance estimates using Cepheids \citep[e.g.][]{kochanek97,kelson00,freedman01,saha06}. A more statistically appropriate method would be to fix the distance scale by introducing a number of distance priors on individual galaxies with distance estimates with Cepheids or Type Ia supernovae or some other method. We do not take this road in order to make our model more transparent.

Another degeneracy comes from defining the reddening zeropoint, applying $E(B-V) \rightarrow E(B-V) + \Delta E(B-V)$ is equivalent to $\overline{M}_i \rightarrow \overline{M}_i - \mathscr{R}_i\Delta E(B-V)$. We remove this degeneracy by fixing the total reddening of SN2012A to $E(B-V) \equiv 0.037$\,mag \citep{tomasella13}. This reddening estimate is based on high-resolution spectroscopy of Na I D lines. The small value of reddening to this supernova guarantees that any potential absolute error in $E(B-V)$ will be relatively small when propagated to other supernovae.

We fix the degeneracy between $\tau$ and $\beta_{n,i}$ by fixing $\alpha_1 \equiv 0.1$ for SN2012aw and $\tau_{\rm thin} \equiv -0.4$ for all supernovae. In principle, it would be sufficient to fix $\tau_{\rm thin}$ only for a single supernova, but we found that this causes artificial features in the model for supernovae with incomplete data around the transition. For the default calculation we fix $\ri$ to the reddening law of \citet{cardelli89} with $\rv \equiv 3.1$. In principle, our model allows for independent determination of the reddening law on a filter-by-filter basis, which we study in Section~\ref{sec:reddening}. 

We impose hard limits on the values of several parameters pertaining to individual supernovae. We constrain $\texp$ to be at least $0.01$\,days before the first observation of every supernova to prevent singularities when calculating $\Pi$. We require that $\omega_1 \ge -1$ so that the photometric radius is increasing with time (Eq.~[\ref{eq:homo}]). We also require that $\omega_2 \ge 0$ to prevent negative expansion velocities and $E(B-V) \ge 0$ to get positive reddenings. If we removed the constraint for the two supernovae with fitted $E(B-V) \equiv 0$, the resulting $E(B-V)$ would be negative only by few thousands of a magnitude.

Finally, we apply a number of priors on individual supernova parameters to aid fitting in cases when there are gaps in photometric coverage or small number or complete lack of expansion velocities. We build these priors based on well-observed supernovae and we take into account any potential correlations between the parameters. We describe the implementation of the priors in the Appendix.


\subsection{Data and fitting method}

In order to constrain the parameters of our model (Eqs.~[\ref{eq:main}--\ref{eq:constraint}]) we require as large a sample of measurements as possible. Our final sample includes $26$ supernovae in $24$ galaxies (there are three supernovae in NGC6946 in our sample). The measurement database consists of roughly $6800$ photometric measurements distributed among $21$ photometric bands with central wavelengths between $0.19$ and $2.2\,\mu$m, and roughly $230$ expansion velocities measured using the Fe II line at $5169$\,\AA. This line is commonly used in the expanding photosphere method \citep[e.g.][]{schmidt94b} so that Equation~(\ref{eq:homo}) is valid. Adding velocities measured on other lines would not necessarily increase the time coverage, because they are often measured from the same spectra. The bulk of the photometric measurements are in $UBVRIJHK$, but we also include the Sloan $griz$ bands, six \swift\ bands $uvw2$, $uvm2$, $uvw1$, and $ubv$, the passband of the ROTSE telescope \citep[SN2006bp;][]{quimby07}, $Z$ band \citep[SN1999em;][]{hamuy01}, and $Y$ band \citep[SN2009N;][]{takats14}. The references we used for obtaining the data are given in Table~\ref{tab:refs}, the photometric bands are summarized in Table~\ref{tab:global}, and the supernovae with corresponding galaxies are listed in Table~\ref{tab:sn_pars}. The master constraint of Equation~(\ref{eq:constraint}) is minimized over $11$ parameters for each supernova minus the three fixed parameters discussed in Section~\ref{sec:priors}, $24$ galaxy distances, and $44$ global parameters, which gives an overall total of $391$ parameters. 

\begin{deluxetable}{lccp{30mm}}
\tabletypesize{\footnotesize}
\tablecolumns{3}
\tablewidth{0pc}
\tablecaption{Database of magnitudes and expansion velocities}
\tablehead{
\colhead{Reference} & \colhead{$N_{\rm phot}$} & \colhead{$N_{\rm vel}$} & Supernovae }
\startdata
\citet{pastorello09} & 354  & 0 & SN2005cs\\
\citet{tomasella13} & 508  & 29 & SN2012A\\
\citet{maguire10b} & 478  & 25 & SN2004et, SN2006my, SN2004A\\
Pastorello (priv. comm) & 0  & 10 & SN2005cs\\
Hamuy (priv. comm) & 464  & 5 & SN1999em\\
\citet{bose13} & 257  & 10 & SN2012aw\\
\citet{inserra12a} & 230  & 11 & SN2009bw\\
\citet{fraser11} & 181  & 6 & SN2009md\\
\citet{roy11} & 171  & 0 & SN2008in\\
\citet{inserra11} & 138  & 7 & SN2007od\\
\citet{gandhi13} & 83  & 1 & SN2009js\\
\citet{hendry05} & 0  & 15 & SN1999em\\
\citet{leonard02a} & 119  & 8 & SN1999gi\\
\citet{vinko06} & 117  & 14 & SN2004dj\\
\citet{gurugubelli08} & 140  & 11 & SN2004A\\
\citet{pozzo06} & 73  & 0 & SN2002hh\\
\citet{munari13} & 408  & 0 & SN2012aw\\
\citet{vandyk12a} & 32  & 0 & SN2008bk\\
\citet{clocchiatti96} & 55  & 5 & SN1992H\\
\citet{pritchard14} & 606  & 0 & SN2005cs, SN2006bp, SN2007od, SN2008in, SN2009dd, SN2009N, SN2012A, SN2012aw\\
\citet{schmidt94a} & 48  & 4 & SN1992am\\
\citet{takats14} & 517  & 17 & SN2009N\\
\citet{pastorello04} & 43  & 0 & SN2001dc\\
\citet{yaron12} & 0  & 6 & SN2001dc\\
\citet{dessart08} & 75  & 0 & SN2006bp\\
\citet{quimby07} & 214  & 9 & SN2006bp\\
\citet{inserra13} & 245  & 22 & SN1996W, SN2009dd, SN2010aj, SN1995ad\\
\citet{dallora14} & 365  & 0 & SN2012aw\\
\citet{barbon82} & 50  & 0 & SN1980K\\
\citet{buta82} & 70  & 0 & SN1980K\\
\citet{uomoto86} & 0  & 3 & SN1980K\\
\citet{korcakova05} & 36  & 0 & SN2004dj\\
\citet{leonard02b} & 200  & 0 & SN1999em\\
\hline
 Total & 6277 & 218

\enddata
\label{tab:refs}
\end{deluxetable}

An improvement over our previous work in \citet{pejcha12} is that we use the Levenberg-Marquardt technique in {\tt cmpfit}\footnote{\url{http://www.physics.wisc.edu/~craigm/idl/cmpfit.html}} \citep{more78,markwardt09} to solve the least-squares problem of Equation~(\ref{eq:constraint}). {\tt cmpfit} can limit each parameter to a specified range, which we use for some parameters (Sec.~\ref{sec:priors}). The implementation of priors in {\tt cmpfit} is described in the Appendix. The non-linear least-squares fitting technique requires partial derivatives of the model with respect to all parameters. In principle, these derivatives can be obtained analytically, but we found that the numerical derivatives calculated in {\tt cmpfit} are more robust.

The complexity of our model precludes obtaining all parameter values from scratch. We started with a small subset of well-observed supernovae with some individual parameters fixed, first obtaining estimates of $\overline{M}_i$ and $\beta_{1,i}$. We then gradually increased the complexity of the model by introducing more photometric bands and the non-linear parts of the model ($\beta_{2,i}$ and $\beta_{3,i}$), alternatively holding either the supernova individual parameters or global parameters fixed. However, in the end, we perform a global fit of all parameters together. This ensures that {\em uncertainties of all parameters are fully coupled and we include covariances between the global parameters of the model and individual supernova parameters}. Our default calculation has $\mathcal{H} \approx 90800$ for $\ndof \approx 6600$ degrees of freedom. 

In this paper, we will discuss also quantities derived from the parameters such as the bolometric luminosities and nickel masses. Since our fit simultaneously adjusts global and individual parameters, collectively denoted as $\mathbf{a}$, we have the full covariance matrix $\mathbf{C}$ of the problem including covariances between global and individual parameters. The uncertainty of a derived quantity $f(\mathbf{a})$ is then
\beq
\sigma_{f}^2 = \left( \frac{\partial f}{\partial \mathbf{a}} \right)^T  \mathbf{C} \left( \frac{\partial f}{\partial \mathbf{a}} \right) = \sum_{i,j} C_{ij}\left(\frac{\partial f}{\partial a_i}\right)\left(\frac{\partial f}{\partial a_j}\right).
\label{eq:uncert_prop}
\eeq
The partial derivatives $\partial f/\partial \mathbf{a}$ are calculated numerically.

\subsection{Relation to previous methods}

Traditionally, the expanding photosphere method requires construction of synthetic magnitudes from theoretical models of supernova spectral evolution. The requisite dilution factors are tabulated for a specific combination of filters ($BV$, $VI$, $BVI$, $JHK$, etc.), where each typically yields a different distance \citep[e.g. Fig.~12 of][]{hamuy01}. Furthermore, this method cannot be used for an arbitrary combination of filters and adding new filters is not straightforward. These shortcomings can be alleviated by modelling of the supernova spectra \citep[e.g.]{dessart06,dessart08,baron07}. In the model presented here, we use the assumption of homologous expansion to construct the evolution of the photospheric radius. We use information from many different supernovae to learn how the photometric color evolves and what are the typical parameters in supernovae and their covariances. Our method naturally works for any combination of filters in our set and adding new filters is relatively easy, requiring a single supernova well-observed in the new band and one of the ``old'' bands.

Similarly, the standardized candle method requires velocity and bolometric measurements at $\texp+50$\,days, which are used as an input to a pre-calibrated correlation \citep{hamuy02}, and this often requires extrapolation to infer quantities at a desired time \citep[e.g. Fig.~3 in][]{olivares10}. We can understand why the standardized candle method works within our model. Evaluating Equation~(\ref{eq:main}) at the moment of $\tau=0$, $t_{\tau=0}$, gives for absolute magnitude $M_i$
\beq
M_i = \overline{M}_i - 5\log v -5\log (t_{\tau=0}-\texp),
\label{eq:scandle}
\eeq
where we assume $w=0$. For our supernovae, $t_{\tau=0}$ typically occurs between $20$ and $50$\,days after $\texp$ with a peak at $35$\,days, making it similar to the usual standardized candle method. Since the chromatic function $\Theta_i=0$ at $\tau=0$, the velocity is a measure of total radius of the supernova if $t_{\tau=0}$ is constant for all supernovae. The fact that $t_{\tau=0}-\texp$ is not constant for all supernovae is responsible for factors $\gtrsim 5$ in front of $\log v$ in the empirical fits \citep{hamuy02,nugent06,poznanski10,olivares10}.

Equation~(\ref{eq:main}) offers another approach to the standardized candle method. Assume that magnitudes $m_i$ and velocity measurements $v$ of a supernova are obtained on a single arbitrary epoch $t$, which is different for each supernova. Equation~(\ref{eq:main}) can then be approximated\footnote{The approximation assumes that the range of observed supernova color indices is narrow, which implies a narrow range of $\tau$. Consequently, the terms proportional to $\tau^2$ and $\tau^3$ in Equation~(\ref{eq:theta}) can be neglected. Note that the $\tau$, $\overline{M}_i$, and $\beta_{1,i}$ can always be redefined so that $\tau=0$ is centered on any desired value of color index.} to provide a fitting formula for the supernova absolute magnitude
\beq
M_i = \overline{M}_i +\ri E(B-V) -K_v\log v - K_t\log(t-\texp) - 2.5\beta_{1,i}\tau.
\label{eq:scm}
\eeq
Here, $\overline{M}_i$, $\ri$, $K_v$, $K_t$, and $\beta_{1,i}$ are fitting coefficients that have a specific meaning within our model (e.g. $K_v=K_t=5$), but their values can be obtained empirically from the data. Equation~(\ref{eq:scm}) reduces to the common standardized candle method by assuming a constant $t-\texp$ and the same color and hence $\tau$ for all supernovae \citep{nugent06,poznanski09,poznanski10,olivares10}. The model in Equation~(\ref{eq:scm}) requires observations in at least three bands per supernova to constrain $\tau$ and $E(B-V)$, and a minimum of $11$ supernovae to constrain the parameters $\overline{M_i}$, $\ri$, $\beta_{1,i}$, $K_v$, and $K_t$ if each supernova is observed in exactly three bands. The requirements on the total number of supernovae can be decreased by assuming a fixed reddening law \citep[e.g.][]{cardelli89} and fitting only for $\rv$. Alternatively, with four bands per supernova, $\rv$ for each supernova can be determined. Number of requisite bands and supernovae can be further decreased by using external values of $E(B-V)$.

To summarize, Equation~(\ref{eq:scm}) differs from the common standardized candle method by not requiring the supernova measurements at any specific post-explosion time and by not assuming the same color for all supernovae. This has practical advantages, because it does not require extrapolation of expansion velocities. This process is uncertain, because each supernova does have a different value of $\omega_1$ as we show in Section~\ref{sec:fits}. Furthermore, uncertainties in determining $\texp$ can be explicitly factored in the fit providing more realistic uncertainties. More degrees of freedom of the model should also provide tighter Hubble diagram. Our sample includes predominantly local objects and is thus not suitable for applying Equation~(\ref{eq:scm}) to construct the Hubble diagram. Applying Equation~(\ref{eq:scm}) to a sample of more distant supernovae will the subject of future work.


\section{Results}
\label{sec:results}

\subsection{Fits of supernovae}
\label{sec:fits}

\begin{figure*}
\centering
\includegraphics[width=0.47\textwidth]{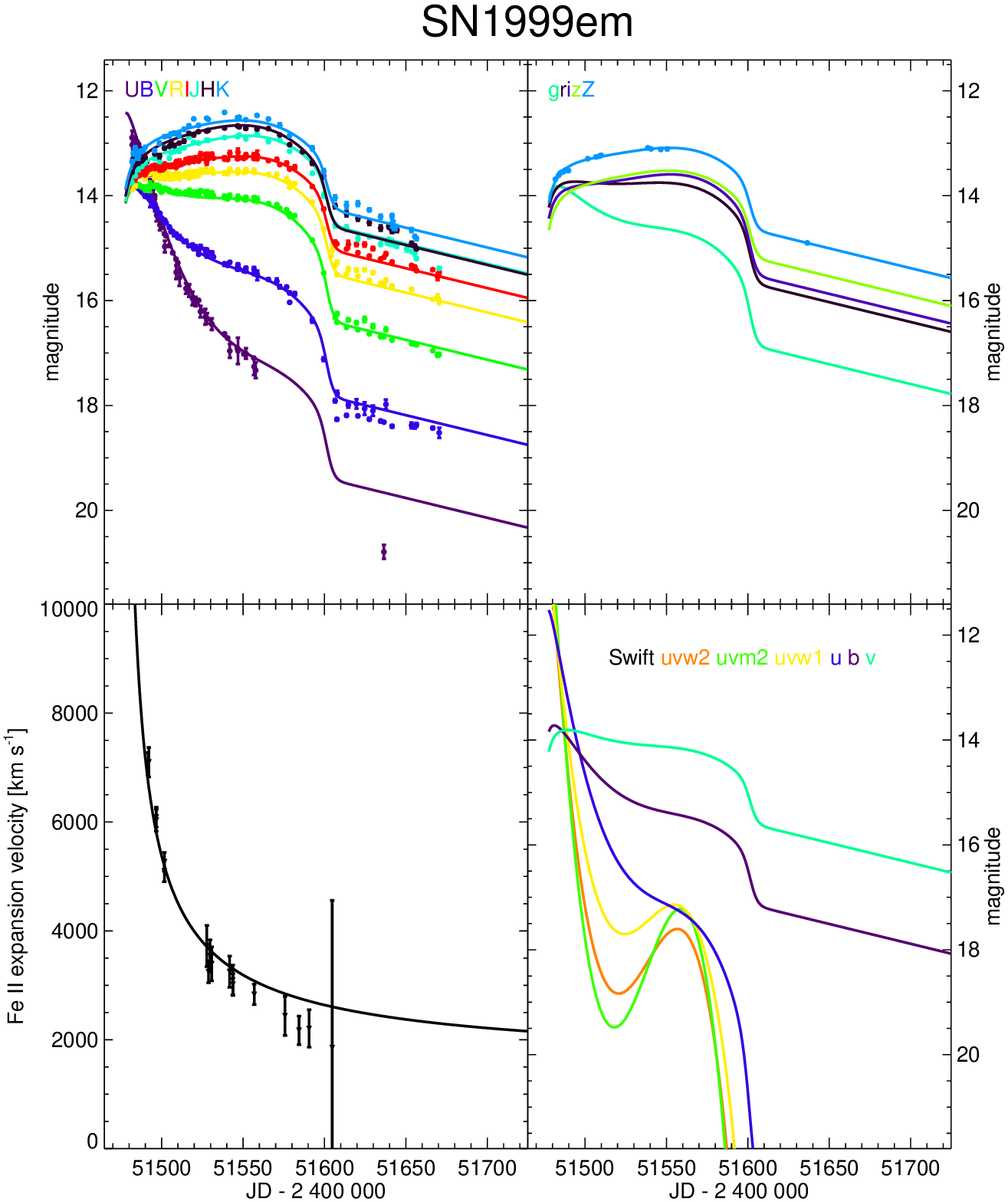}
\includegraphics[width=0.47\textwidth]{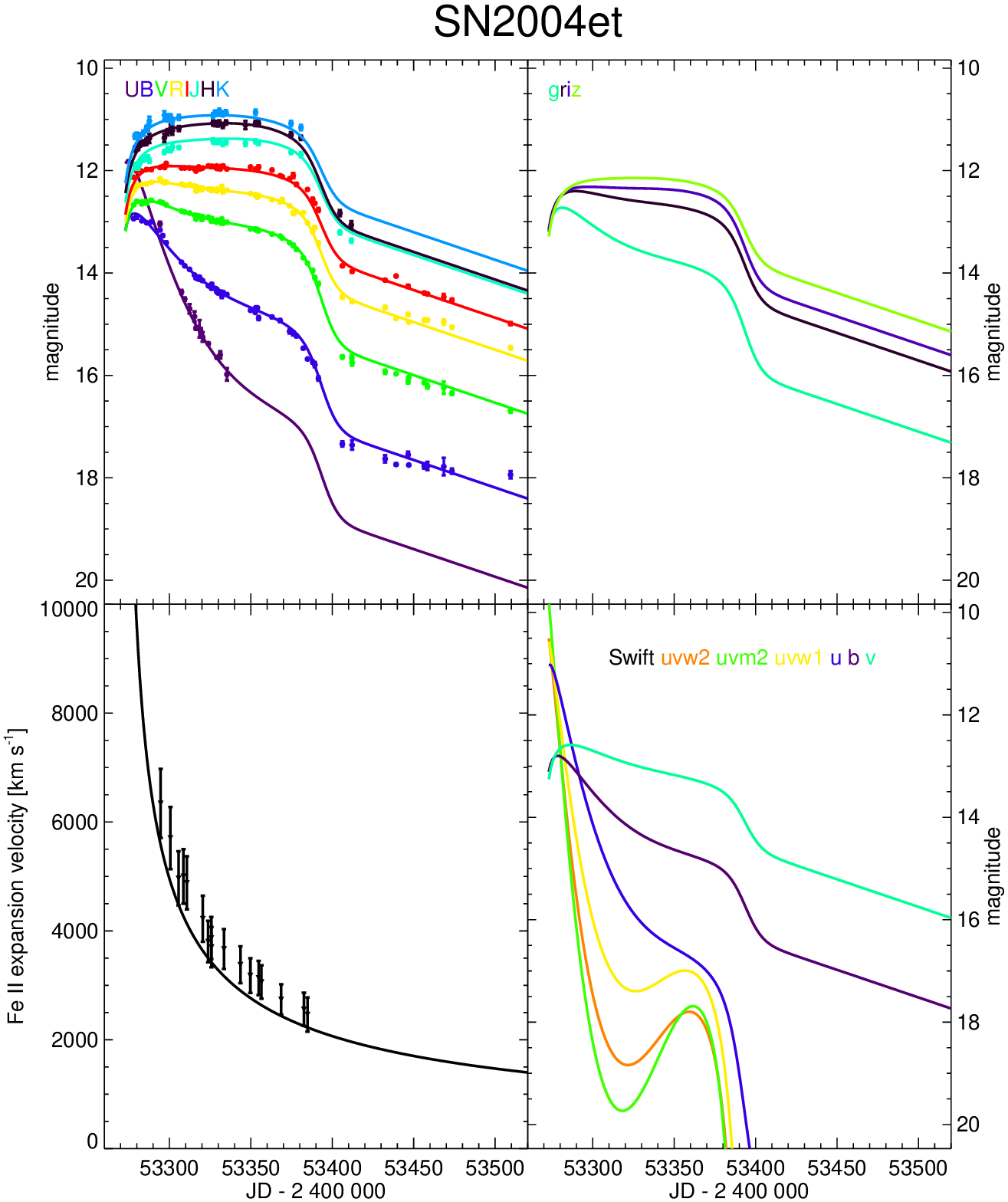}
\includegraphics[width=0.47\textwidth]{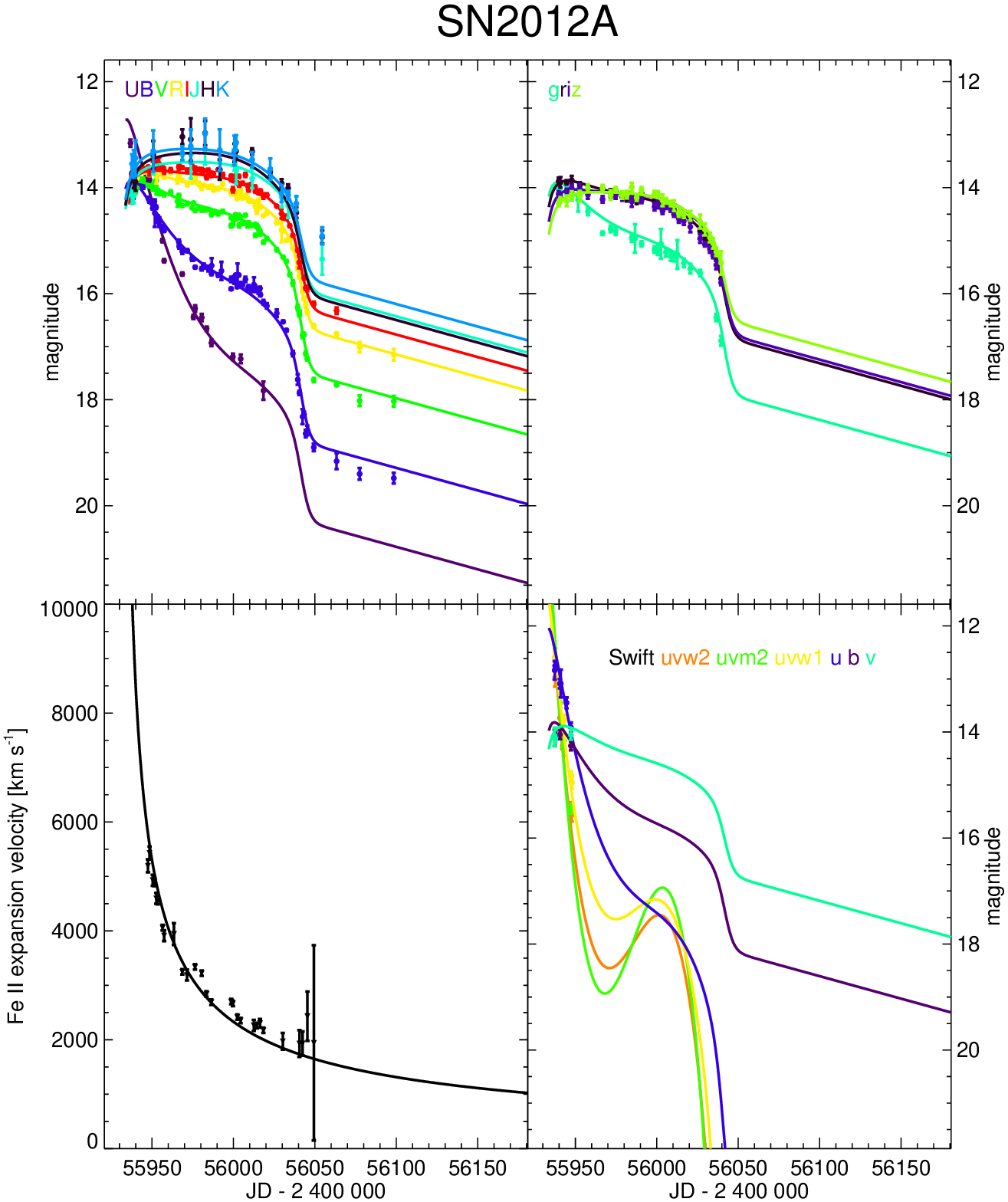}
\includegraphics[width=0.47\textwidth]{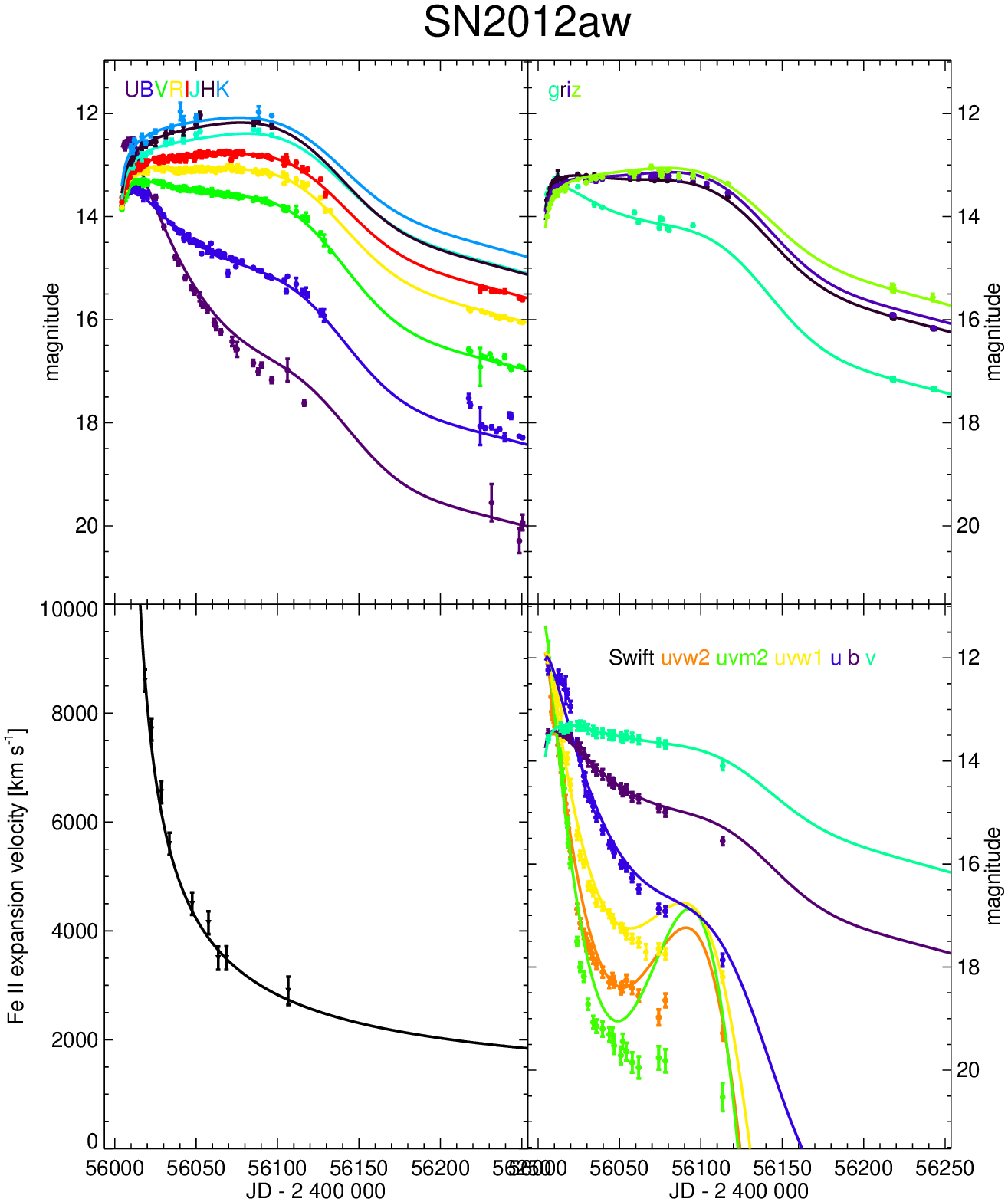}
\caption{An illustration of our fits to the light curves and expansion velocities of several well-known supernovae: SN1999em (top left), SN2004et (top right), SN2012A (bottom left), and SN2012aw (bottom right). The uncertainties of expansion velocities are divided by $1-w$ to reflect their weight in the fitting (Eq.~[\ref{eq:constraint}]). The fits to the remaining objects in our sample are available in the electron edition of the journal.}
\label{fig:lc}
\end{figure*}

\begin{deluxetable*}{lllccccccc}
\tabletypesize{\footnotesize}
\tablecolumns{9}
\tablewidth{0pc}
\tablecaption{Table of results}
\tablehead{\colhead{Supernova} & \colhead{Galaxy} & \colhead{$\texp$} & \colhead{$\tp$ [d]} & \colhead{$\tw$ [d]} & \colhead{$E(B-V)$} &  \colhead{$N_{\rm phot}$} & \colhead{$N_{\rm vel}$} & \colhead{$\chi^2$}}
\startdata
SN1980K    & NGC6946              & $ 44539.36 \pm  0.00 $ & $  60.95 \pm 1.14 $ & $ 12.62 \pm 0.48 $ & $  0.000 \pm 0.000 $ &  120 &    3 &   234.7 \\
SN1992am   & MCG-01-04-039        & $ 48806.67 \pm  2.21 $ & $ 119.78 \pm 5.86 $ & $  4.39 \pm 3.00 $ & $  0.218 \pm 0.015 $ &   48 &    4 &    38.2 \\
SN1992H    & NGC5377              & $ 48650.16 \pm  2.06 $ & $ 128.53 \pm 3.37 $ & $ 19.96 \pm 1.02 $ & $  0.000 \pm 0.000 $ &   55 &    5 &  1767.2 \\
SN1995ad   & NGC2139              & $ 49969.33 \pm  3.81 $ & $  84.62 \pm 9.61 $ & $  2.90 \pm 3.31 $ & $  0.403 \pm 0.009 $ &   52 &    9 &   208.8 \\
SN1996W    & NGC4027              & $ 50180.13 \pm  1.26 $ & $  99.26 \pm 12.62 $ & $  6.91 \pm 5.28 $ & $  0.329 \pm 0.011 $ &   59 &    5 &   677.0 \\
SN1999em   & NGC1637              & $ 51474.80 \pm  0.36 $ & $ 126.87 \pm 0.38 $ & $  2.32 \pm 0.08 $ & $  0.161 \pm 0.002 $ &  664 &   20 & 10037.2 \\
SN1999gi   & NGC3184              & $ 51521.62 \pm  0.14 $ & $ 122.87 \pm 0.49 $ & $  5.19 \pm 0.27 $ & $  0.376 \pm 0.006 $ &  119 &    8 &   674.5 \\
SN2001dc   & NGC5777              & $ 52049.02 \pm  4.97 $ & $ 117.13 \pm 4.96 $ & $  0.10 \pm 0.00 $ & $  0.678 \pm 0.015 $ &   43 &    6 &    36.4 \\
SN2002hh   & NGC6946              & $ 52574.86 \pm  4.32 $ & $ 139.93 \pm 10.81 $ & $  0.11 \pm 4.41 $ & $  1.565 \pm 0.006 $ &   73 &    0 &  1690.7 \\
SN2004A    & NGC6207              & $ 53001.53 \pm  0.77 $ & $ 123.28 \pm 0.82 $ & $  3.90 \pm 0.09 $ & $  0.101 \pm 0.003 $ &  153 &   11 &  3828.9 \\
SN2004dj   & NGC2403              & $ 53191.45 \pm  2.90 $ & $ 103.30 \pm 2.87 $ & $  2.72 \pm 0.37 $ & $  0.012 \pm 0.006 $ &  153 &   14 &   602.3 \\
SN2004et   & NGC6946              & $ 53270.03 \pm  0.16 $ & $ 122.40 \pm 0.31 $ & $  5.92 \pm 0.19 $ & $  0.384 \pm 0.003 $ &  401 &   19 & 11484.4 \\
SN2005cs   & M51~(NGC5194)        & $ 53546.78 \pm  0.71 $ & $ 127.57 \pm 0.71 $ & $  1.38 \pm 0.06 $ & $  0.187 \pm 0.003 $ &  411 &   10 &  9743.4 \\
SN2006bp   & NGC3953              & $ 53834.60 \pm  0.14 $ & $  86.25 \pm 0.68 $ & $  4.67 \pm 0.23 $ & $  0.422 \pm 0.006 $ &  394 &    9 &  2243.2 \\
SN2006my   & NGC4651              & $ 53995.28 \pm  5.97 $ & $  95.58 \pm 5.90 $ & $  1.24 \pm 0.29 $ & $  0.048 \pm 0.006 $ &   64 &    6 &   721.1 \\
SN2007od   & UGC12846             & $ 54379.12 \pm  2.01 $ & $ 131.67 \pm 1.90 $ & $ 16.60 \pm 0.41 $ & $  0.107 \pm 0.004 $ &  195 &    7 &  3171.4 \\
SN2008bk   & NGC7793              & $ 54548.18 \pm  4.78 $ & $ 127.18 \pm 7.49 $ & $  4.00 \pm 2.18 $ & $  0.109 \pm 0.016 $ &   32 &    0 &   186.2 \\
SN2008in   & M61~(NGC4303)        & $ 54816.12 \pm  0.34 $ & $ 117.47 \pm 0.35 $ & $  1.66 \pm 0.07 $ & $  0.162 \pm 0.003 $ &  212 &    0 &  8435.0 \\
SN2009bw   & UGC02890             & $ 54918.22 \pm  0.02 $ & $ 132.88 \pm 0.35 $ & $  1.18 \pm 0.20 $ & $  0.432 \pm 0.004 $ &  230 &   11 &  4077.1 \\
SN2009dd   & NGC4088              & $ 54928.13 \pm  1.31 $ & $ 124.37 \pm 1.76 $ & $  5.75 \pm 0.79 $ & $  0.367 \pm 0.007 $ &   94 &    4 &  1377.2 \\
SN2009js   & NGC0918              & $ 55097.85 \pm  2.85 $ & $ 127.10 \pm 3.09 $ & $  2.84 \pm 0.83 $ & $  0.475 \pm 0.014 $ &   83 &    1 &    53.2 \\
SN2009md   & NGC3389              & $ 55163.28 \pm  0.95 $ & $ 116.90 \pm 0.94 $ & $  1.81 \pm 0.08 $ & $  0.201 \pm 0.007 $ &  181 &    6 &   625.4 \\
SN2009N    & NGC4487              & $ 54840.55 \pm  0.30 $ & $ 116.64 \pm 0.31 $ & $  1.15 \pm 0.04 $ & $  0.307 \pm 0.002 $ &  547 &   17 &  7805.2 \\
SN2010aj   & MCG-01-32-035        & $ 55260.48 \pm  4.32 $ & $  87.91 \pm 4.28 $ & $  4.10 \pm 0.26 $ & $  0.181 \pm 0.025 $ &   54 &    4 &   194.4 \\
SN2012A    & NGC3239              & $ 55930.72 \pm  0.13 $ & $ 111.12 \pm 0.17 $ & $  2.87 \pm 0.09 $ & $ \equiv  0.037 $ &  538 &   29 &  5179.3 \\
SN2012aw   & M95~(NGC3351)        & $ 56003.63 \pm  0.12 $ & $ 136.04 \pm 1.11 $ & $ 27.27 \pm 0.52 $ & $  0.173 \pm 0.002 $ & 1302 &   10 & 11359.1

\enddata
\label{tab:sn_pars}
\end{deluxetable*}

\begin{deluxetable*}{cccccccc}
\tabletypesize{\footnotesize}
\tablecolumns{11}
\tablewidth{0pc}
\tablecaption{Table of model specific parameters}
\tablehead{\colhead{Supernova} & \colhead{$10^{3}\alpha_0$} & \colhead{$\alpha_1$}  &  \colhead{$\omega_0$ [$10^3$\,\kms]} & \colhead{$\omega_1$} & \colhead{$\omega_2$  [$10^3$\,\kms]} & \colhead{$10^{3}\gamma_0$} & \colhead{$\gamma_1$}}
\startdata
SN1980K    &  $ -1.63 \pm 0.19 $ & $  0.0753 \pm 0.0014 $ & $ 110.4 \pm  3.1 $ & $  -1.000 \pm  0.000 $ & $  0.00 \pm 0.00 $ & $   -3.47 \pm   0.11 $ & $   10.34 \pm   0.03 $ \\
SN1992am   &  $ -3.36 \pm 0.15 $ & $  0.1326 \pm 0.0057 $ & $  57.7 \pm  5.4 $ & $  -0.596 \pm  0.022 $ & $  0.00 \pm 0.00 $ & $   -4.18 \pm   0.49 $ & $   11.64 \pm   0.07 $ \\
SN1992H    &  $ -3.61 \pm 0.04 $ & $  0.1495 \pm 0.0069 $ & $  73.4 \pm  6.3 $ & $  -0.632 \pm  0.025 $ & $  0.00 \pm 0.00 $ & $   -4.19 \pm   0.04 $ & $   11.86 \pm   0.06 $ \\
SN1995ad   &  $ -1.84 \pm 0.11 $ & $  0.1457 \pm 0.0067 $ & $  68.6 \pm 10.2 $ & $  -0.741 \pm  0.056 $ & $  0.06 \pm 0.33 $ & $   -4.34 \pm   0.14 $ & $   11.54 \pm   0.10 $ \\
SN1996W    &  $ -2.51 \pm 0.06 $ & $  0.0906 \pm 0.0035 $ & $  68.8 \pm  3.5 $ & $  -0.699 \pm  0.019 $ & $  0.00 \pm 0.00 $ & $   -3.57 \pm   0.24 $ & $   11.49 \pm   0.07 $ \\
SN1999em   &  $ -3.74 \pm 0.01 $ & $  0.1206 \pm 0.0014 $ & $  42.9 \pm  1.4 $ & $  -0.740 \pm  0.013 $ & $  1.44 \pm 0.04 $ & $   -3.03 \pm   0.03 $ & $   10.99 \pm   0.02 $ \\
SN1999gi   &  $ -3.30 \pm 0.04 $ & $  0.1104 \pm 0.0012 $ & $  50.5 \pm  1.6 $ & $  -0.819 \pm  0.010 $ & $  1.02 \pm 0.04 $ & $   -4.27 \pm   0.25 $ & $   10.86 \pm   0.02 $ \\
SN2001dc   &  $ -3.98 \pm 0.16 $ & $  0.1081 \pm 0.0170 $ & $  11.0 \pm  2.9 $ & $  -0.585 \pm  0.103 $ & $  0.32 \pm 0.21 $ & $   -3.41 \pm   0.52 $ & $   10.01 \pm   0.08 $ \\
SN2002hh   &  $ -1.42 \pm 0.10 $ & $ -0.1151 \pm 0.0139 $ & $ 185.7 \pm 10.1 $ & $  -0.950 \pm  0.015 $ & $  0.00 \pm 0.00 $ & $   -5.22 \pm   0.01 $ & $   11.30 \pm   0.06 $ \\
SN2004A    &  $ -3.32 \pm 0.03 $ & $  0.1106 \pm 0.0031 $ & $  20.7 \pm  1.2 $ & $  -0.474 \pm  0.012 $ & $  0.00 \pm 0.00 $ & $   -3.66 \pm   0.02 $ & $   11.02 \pm   0.04 $ \\
SN2004dj   &  $ -4.46 \pm 0.12 $ & $  0.1078 \pm 0.0125 $ & $  31.3 \pm  3.1 $ & $  -0.584 \pm  0.028 $ & $  0.00 \pm 0.00 $ & $   -2.54 \pm   0.03 $ & $   10.50 \pm   0.05 $ \\
SN2004et   &  $ -3.37 \pm 0.02 $ & $  0.1185 \pm 0.0008 $ & $  37.7 \pm  1.2 $ & $  -0.596 \pm  0.006 $ & $  0.00 \pm 0.00 $ & $   -4.31 \pm   0.01 $ & $   10.96 \pm   0.03 $ \\
SN2005cs   &  $ -3.66 \pm 0.01 $ & $  0.1106 \pm 0.0026 $ & $  42.1 \pm  1.2 $ & $  -1.000 \pm  0.000 $ & $  1.33 \pm 0.06 $ & $   -3.14 \pm   0.04 $ & $    9.97 \pm   0.04 $ \\
SN2006bp   &  $ -2.85 \pm 0.03 $ & $  0.0994 \pm 0.0006 $ & $  81.7 \pm  2.9 $ & $  -0.813 \pm  0.009 $ & $  0.73 \pm 0.07 $ & $   -2.70 \pm   0.13 $ & $    9.56 \pm   0.06 $ \\
SN2006my   &  $ -4.75 \pm 0.17 $ & $  0.0591 \pm 0.0247 $ & $   9.0 \pm  2.7 $ & $  -0.373 \pm  0.104 $ & $  0.18 \pm 0.37 $ & $   -3.34 \pm   0.08 $ & $   10.32 \pm   0.09 $ \\
SN2007od   &  $ -2.17 \pm 0.03 $ & $  0.1327 \pm 0.0042 $ & $  74.4 \pm  3.0 $ & $  -0.695 \pm  0.014 $ & $  0.00 \pm 0.00 $ & $   -5.10 \pm   0.13 $ & $   10.31 \pm   0.04 $ \\
SN2008bk   &  $ -3.08 \pm 0.42 $ & $  0.1035 \pm 0.0134 $ & $  35.4 \pm 13.4 $ & $  -0.807 \pm  0.108 $ & $  0.92 \pm 0.23 $ & $   -4.41 \pm   0.03 $ & $   10.60 \pm   0.21 $ \\
SN2008in   &  $ -3.78 \pm 0.02 $ & $  0.1339 \pm 0.0013 $ & $  28.6 \pm  8.5 $ & $  -0.548 \pm  0.007 $ & $  0.00 \pm 0.00 $ & $   -3.54 \pm   0.04 $ & $   10.86 \pm   0.26 $ \\
SN2009bw   &  $ -2.86 \pm 0.02 $ & $  0.1053 \pm 0.0004 $ & $  72.8 \pm  1.2 $ & $  -0.807 \pm  0.003 $ & $  0.23 \pm 0.02 $ & $   -5.41 \pm   0.18 $ & $   10.62 \pm   0.02 $ \\
SN2009dd   &  $ -3.68 \pm 0.06 $ & $  0.1053 \pm 0.0048 $ & $  49.2 \pm  2.9 $ & $  -0.767 \pm  0.034 $ & $  0.39 \pm 0.09 $ & $   -4.03 \pm   0.13 $ & $   10.63 \pm   0.12 $ \\
SN2009js   &  $ -3.54 \pm 0.08 $ & $  0.1281 \pm 0.0098 $ & $  34.9 \pm  4.8 $ & $  -0.489 \pm  0.036 $ & $  0.00 \pm 0.00 $ & $   -4.47 \pm   0.27 $ & $   11.20 \pm   0.12 $ \\
SN2009md   &  $ -3.54 \pm 0.06 $ & $  0.1043 \pm 0.0040 $ & $  22.1 \pm  1.6 $ & $  -0.592 \pm  0.039 $ & $  0.06 \pm 0.15 $ & $   -3.78 \pm   0.20 $ & $   10.29 \pm   0.06 $ \\
SN2009N    &  $ -4.21 \pm 0.01 $ & $  0.1551 \pm 0.0013 $ & $  10.8 \pm  0.3 $ & $  -0.332 \pm  0.007 $ & $  0.00 \pm 0.00 $ & $   -3.85 \pm   0.02 $ & $   10.86 \pm   0.02 $ \\
SN2010aj   &  $ -2.14 \pm 0.13 $ & $  0.0978 \pm 0.0101 $ & $ 113.9 \pm  8.5 $ & $  -0.851 \pm  0.027 $ & $  0.00 \pm 0.00 $ & $   -4.10 \pm   0.54 $ & $   10.58 \pm   0.09 $ \\
SN2012A    &  $ -4.12 \pm 0.02 $ & $  0.1099 \pm 0.0007 $ & $  35.4 \pm  0.5 $ & $  -0.642 \pm  0.004 $ & $  0.00 \pm 0.00 $ & $   -3.39 \pm   0.04 $ & $   10.36 \pm   0.02 $ \\
SN2012aw   &  $ -2.85 \pm 0.02 $ & $ \equiv 0.1000 $ &   $  58.0 \pm  1.9 $ & $  -0.744 \pm  0.011 $ & $  0.89 \pm 0.07 $ & $   -3.70 \pm   0.05 $ & $   11.17 \pm   0.03 $ 

\enddata
\label{tab:sn_pars2}
\end{deluxetable*}

In Figure~\ref{fig:lc} we show the resulting fits to the light curves and expansion velocities for several well-observed supernovae. The fits of the remaining objects are available in the on-line version of the journal. The behavior of the model in supernovae with less data than what is shown in Figure~\ref{fig:lc} can be ascertained later in Figure~\ref{fig:sn1980k} or by the fits of ASAS13co in \citet{holoien14}.  Fit parameters for individual supernovae are given in Tables~\ref{tab:sn_pars} and \ref{tab:sn_pars2}. We see that our model reproduces well the key features of Type II-P supernova light curves and can fit them simultaneously in many bands. The model reproduces the light curve just after the explosion when the optical flux rises due to photospheric expansion and cooling so that peak of the spectral energy distribution moves into the optical. At the same time, the supernova color evolves from blue to red, which is captured by the rapidly decreasing flux in the $U$ band and in the \swift\ near-UV bands. Another key feature is the plateau, where our model reproduces the observed variety between supernovae, ranging from the rather flat plateau of SN1999em to the gradually declining one in SN2012A. Furthermore, our model naturally reproduces the flattening of the decline in the $U$ and $B$ bands just before the plateau ends. Our model predicts that this flattening should manifest itself as a distinct bump in the near-UV \swift\ bands. The only supernova with \swift\ data covering this part of the light curve is SN2012aw and we indeed see indications of a marked flattening and possibly upturn of brightness in $uvm2$. However, the data are missing just during the predicted bump and the supernova was already very faint. This part of the near-UV evolution could be better constrained with more data, which could allow higher order determination of the temperature coefficients (Eq.~[\ref{eq:theta}]).

Our model does reasonably well in reproducing the transition from the optically-thick plateau to the optically-thin exponential decay part of the light curve. We fit very sharp transitions such as in SN2005cs as well as more gradual transitions such as in SN2004et. In objects without observations during the transition phase like SN2012aw, the model typically favors very long $\tw$. This does not affect the quantities derived from the model, because $\tw$ has little correlation with other parameters as we show below. We also see that our assumption of constant $\tau_{\rm thick} \equiv -0.4$ and hence constant color during the exponential decay is reasonably good in objects like SN2012A and SN2012aw. In some objects like SN2004et or SN2005cs we observe a distinct color evolution during this phase. However, in this paper we focus predominantly on the supernova plateau, which exhibits the highest luminosities and allows for distance determination. Improving the description of the exponential decay phase is a subject for future work.

In Figure~\ref{fig:lc}, we also show fits to the expansion velocities, which show excellent agreement with the data. In the case of SN2004et, we observe a systematic shift between the data and the model. The reason is that the distance to SN2004et host galaxy NGC6946 is constrained also by the data of SN2002hh and SN1980K, which results in a small systematic offset in expansion velocities of SN2004et. If SN2004et was fit independently, then there would be no systematic shift between the observed expansion velocities and the model.  In several supernovae like SN1999em, the fit deviates from the observations close to $\tw$. This probably indicates that these velocities are biased because the supernova is becoming transparent. We emphasize that in our model, the radius changes in supernovae with incomplete velocity coverage are constrained not only by velocities but also by the achromatic part of the photometry.

Now we investigate the range of validity of our model. Both the total value of $\mathcal{H}$ and the $\chi^2$ values of individual objects given in Table~\ref{tab:sn_pars} indicate that $\chi^2/\ndof \sim 5$ to $10$. Normally, this would indicate that the model is not fitting the data well and it should be modified. However, our model is phenomenological and we would not expect it to provide a perfect match to data of all supernovae. The agreement between the data and the model could be improved by adding degrees of freedom in the model, for example, by using higher-order temperature terms in Equation~(\ref{eq:theta}) or by assuming non-linear evolution of $\tau$ in Equation~(\ref{eq:tauthick}). However, we prefer to keep the model as simple as possible, even at the cost of higher $\mathcal{H}/\ndof$. One way to rectify this situation is to multiply the measurement uncertainties by $\sqrt{\mathcal{H}/\ndof}$ and repeat the fit, which ensures that $\mathcal{H}/\ndof \equiv 1$. At several places of the paper, we apply this procedure to obtain perhaps more realistic uncertainties of the fit parameters and we explicitly mention when we do so. Uncertainties in Tables~\ref{tab:sn_pars} and \ref{tab:sn_pars2} are not rescaled.

Another aspect of the phenomenological description of supernova light curves and expansion velocities is the possibility to arrive to misleading results if the model is pushed too far. For example, assume that the ``true'' evolution of $\tau$ in Equation~(\ref{eq:tauthick}) is nonlinear. If there are no observations of the initial part of the light curve, the two coefficients in Equation~(\ref{eq:tauthick}) will be different than if the full light curve was fitted. As a result, $\tau$ in the initial part of the light curve will be wrong, which can have dramatic consequences, because of the temperature sensitivity of the blue bands (Fig.~\ref{fig:koef}). Therefore, we will present bolometric light curves and other quantities only for epochs after the first observation of each supernova.

\begin{figure}
\plotone{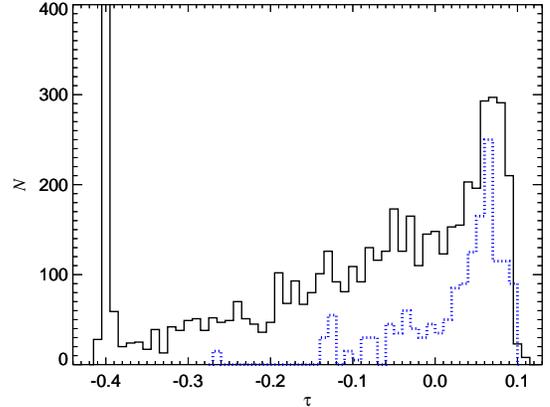}
\caption{Distribution of $\tau$ assigned to the photometric measurements within our model (solid black line). We also overplot a histogram for the \swift\ $uvw2$, $uvm2$, and $uvw1$ bands (blue dotted line, multiplied by a factor $5$ for the sake of clarity). For the sake of clarity, we do not show the full height of the peak at $\tau = -0.4$.}
\label{fig:hist_temp}
\end{figure}

Despite this caveat, if $\tau$ is well determined by observations in a subset of bands, we can robustly predict the flux in the remaining bands if the coefficients $\beta_{n,i}$ are well-constrained at this $\tau$ by observations of some other supernova. In Figure~\ref{fig:hist_temp} we present the distribution of $\tau$ assigned to all our photometric measurements by the best-fit model. We see that there are plenty of observations for $-0.4 \le \tau \le 0.11$ suggesting that there are no global ``holes'' in our coverage of $\tau$. The peak at $\tau = -0.4$ is due to the fact that we fix $\tau_{\rm thin} \equiv -0.4$ during the exponential decay part of the light curve. We will show in Section~\ref{sec:sed} that the near-UV \swift\ bands are extremely important to constrain the total flux from supernovae in the early part of the light curve. We show in Figure~\ref{fig:hist_temp} with dotted blue line the distribution of the three bluest \swift\ bands. We see that we can reliably predict near-UV magnitudes only for $\tau \le 0.095$. There is a lack of near-UV measurements at low $\tau$ as well, but this is not important for bolometric fluxes due to the negligible contribution of the near-UV to the flux in later epochs. However, we find that the coefficients $\beta_{n,i}$ are only determined reliably for $\tau \ge -0.06$ for the \swift\ bands.

\begin{figure*}
\plotone{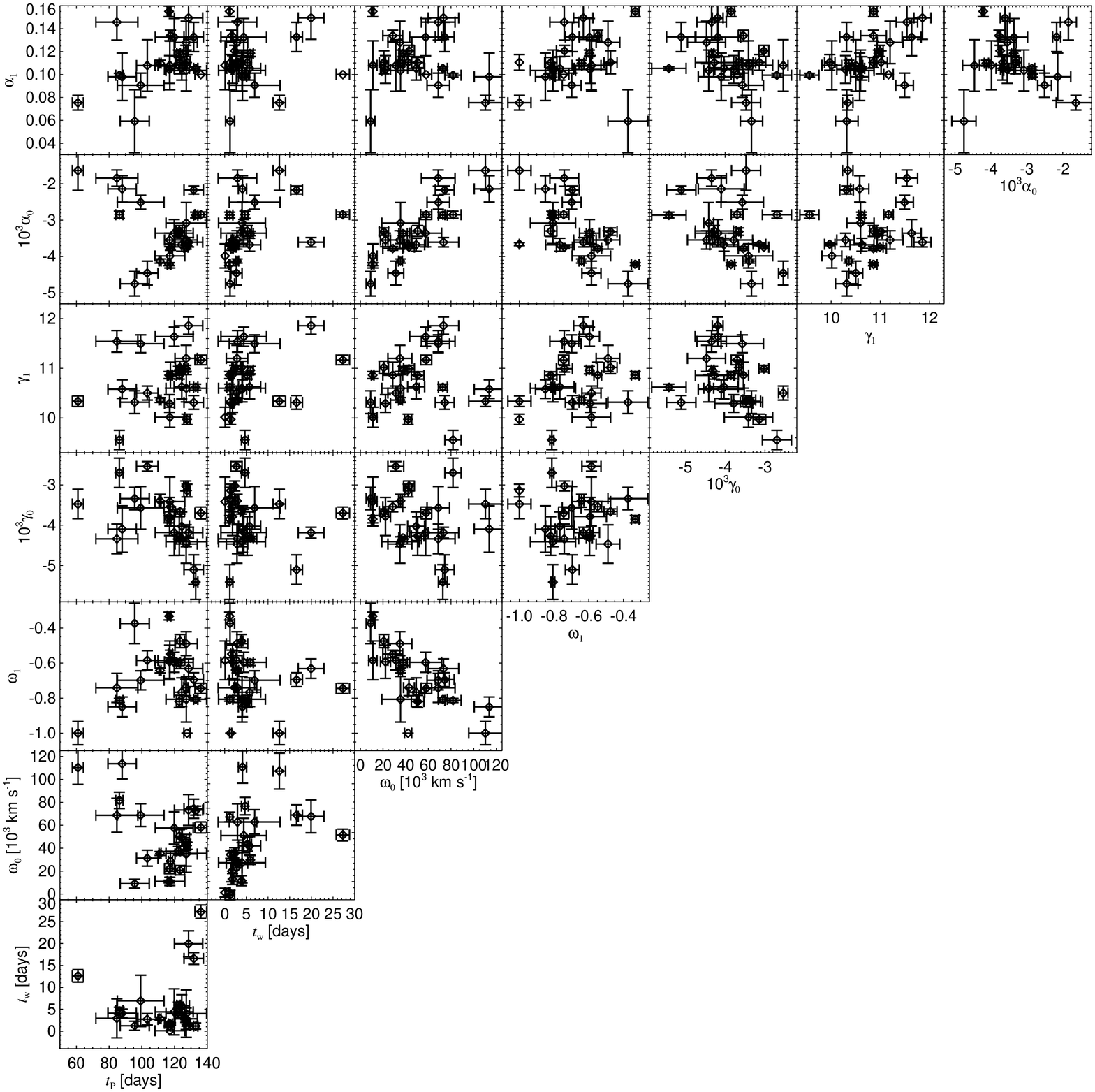}
\caption{Correlations of the individual supernova parameters $\tp$, $\tw$, $\omega_0$, $\omega_1$, $\gamma_0$, $\gamma_1$, $\alpha_0$, and $\alpha_1$. The uncertainties come from a modified fit with $\mathcal{H}/\ndof = 1$.}
\label{fig:cor}
\end{figure*}

In Figure~\ref{fig:cor} we show the mutual dependencies of the individual supernova parameters from Tables~\ref{tab:sn_pars} and \ref{tab:sn_pars2}. The uncertainties come from a modified fit with $\mathcal{H}/\ndof =1$. We see that the plateau duration $\tp$ correlates with almost all parameters except for several outliers with $\tp\lesssim 90$\,d, which typically have poor coverage of the early phases of the light curve. We also see that $\omega_0$ and $\omega_1$ (Eq.~[\ref{eq:v}]) correlate in the sense that velocity decays faster in supernovae with initially higher velocity. Both of these parameters correlate with $\alpha_0$, which is the slope of the temperature parameter decrease (Eq.~[\ref{eq:tauthick}]), in the sense that in supernovae with initially higher expansion velocity the temperature parameter decays more slowly. Our values of $\omega_1$ are typically smaller than the mean obtained by \citet{nugent06} and \citet{faran14}. The reason is that we have also a constant velocity offset $\omega_2 \ge 0$ in our model, which typically makes $\omega_1$ more negative. We have a different value of $\omega_1$ for each of our supernova and $\omega_1$ is determined not only by velocity measurements, but also by the photometry.

\begin{deluxetable*}{cclcccccc}
\tabletypesize{\footnotesize}
\tablecaption{Global parameters}
\tablecolumns{9}
\tablewidth{0pc}
\tablehead{\colhead{Filter} & \colhead{$\lambda$ [$\mu$m]} & \colhead{$N$} & \colhead{$\mathcal{F}_{0}$ [erg\,cm$^{-2}$\,s$^{-1}$\,\AA$^{-1}$] } & \colhead{$\overline{M}$} & \colhead{$\beta_{1}$} & \colhead{$\beta_{2}$} & \colhead{$\beta_{3}$}}
\startdata
$U$  &   0.36 & \phn 268    &  $ 3.98 \times 10^{ -9}$ &  $ 10.089 \pm 0.048  $  & $ 11.900 \pm 0.037  $  &\phn $   78.43 \pm   0.54  $  & $    363.0 \pm     4.7  $  &\\
$B$  &   0.44 & \phn 971    &  $ 6.95 \times 10^{ -9}$ & \phn $  9.801 \pm 0.045  $  &\phn $  6.976 \pm 0.032  $  &\phn $   46.62 \pm   0.29  $  & $    265.0 \pm     2.3  $  &\\
$V$  &   0.55 &  1159   &  $ 3.63 \times 10^{ -9}$ & \phn $  9.240 \pm 0.043  $  &\phn $  4.309 \pm 0.032  $  &\phn $   31.11 \pm   0.26  $  & $    221.9 \pm     1.9  $  &\\
$R$  &   0.66 &  1119   &  $ 2.25 \times 10^{ -9}$ & \phn $  8.972 \pm 0.043  $  &\phn $  3.578 \pm 0.032  $  &\phn $   24.76 \pm   0.26  $  & $    181.6 \pm     2.0  $  &\\
$I$  &   0.81 & \phn 987    &  $ 1.20 \times 10^{ -9}$ & \phn $  8.890 \pm 0.042  $  &\phn $  3.072 \pm 0.032  $  &\phn $   20.20 \pm   0.26  $  & $    156.3 \pm     2.0  $  &\\
$J$  &   1.22 & \phn 175    &  $ 3.40 \times 10^{-10}$ & \phn $  8.791 \pm 0.042  $  &\phn $  2.462 \pm 0.034  $  &\phn $   15.76 \pm   0.35  $  & $    138.1 \pm     2.4  $  &\\
$H$  &   1.63 & \phn 180    &  $ 1.26 \times 10^{-10}$ & \phn $  8.651 \pm 0.042  $  &\phn $  2.258 \pm 0.035  $  &\phn $   12.94 \pm   0.38  $  & $    133.0 \pm     2.6  $  &\\
$K$  &   2.21 & \phn 138    &  $ 3.90 \times 10^{-11}$ & \phn $  8.563 \pm 0.042  $  &\phn $  2.366 \pm 0.036  $  &\phn $   12.49 \pm   0.40  $  & $    117.8 \pm     2.7  $  &\\
$g$  &   0.49 & \phn\phn 83     &  $ 4.49 \times 10^{ -9}$ & \phn $  9.482 \pm 0.044  $  &\phn $  5.442 \pm 0.038  $  &\phn $   38.37 \pm   0.51  $  & $    239.5 \pm     3.5  $  &\\
$r$  &   0.63 & \phn 120    &  $ 2.71 \times 10^{ -9}$ & \phn $  9.127 \pm 0.043  $  &\phn $  3.730 \pm 0.035  $  &\phn $   25.77 \pm   0.45  $  & $    183.7 \pm     3.1  $  &\\
$i$  &   0.78 & \phn 116    &  $ 1.79 \times 10^{ -9}$ & \phn $  9.252 \pm 0.043  $  &\phn $  3.122 \pm 0.034  $  &\phn $   23.30 \pm   0.38  $  & $    181.7 \pm     2.6  $  &\\
$z$  &   0.93 & \phn\phn 98     &  $ 1.26 \times 10^{ -9}$ & \phn $  9.312 \pm 0.042  $  &\phn $  2.670 \pm 0.035  $  &\phn $   16.46 \pm   0.42  $  & $    136.0 \pm     3.1  $  &\\
Swift $uvw2$  &   0.19 & \phn 105    &  $ 5.23 \times 10^{ -9}$ &  $ 12.673 \pm 0.056  $  & $ 14.814 \pm 0.130  $  & $  300.73 \pm   3.03  $  & $   2478.3 \pm    45.6  $  &\\
Swift $uvm2$  &   0.22 & \phn 101    &  $ 4.01 \times 10^{ -9}$ &  $ 13.212 \pm 0.059  $  & $ 15.592 \pm 0.175  $  & $  376.97 \pm   3.62  $  & $   3163.2 \pm    64.2  $  &\\
Swift $uvw1$  &   0.26 & \phn 105    &  $ 4.26 \times 10^{ -9}$ &  $ 11.704 \pm 0.053  $  & $ 13.742 \pm 0.101  $  & $  233.78 \pm   2.70  $  & $   1865.0 \pm    37.6  $  &\\
Swift $u$  &   0.35 & \phn 108    &  $ 3.25 \times 10^{ -9}$ &  $ 10.385 \pm 0.051  $  & $ 13.301 \pm 0.082  $  & $  120.23 \pm   2.53  $  & $    733.2 \pm    35.7  $  &\\
Swift $b$  &   0.44 & \phn\phn 99     &  $ 5.82 \times 10^{ -9}$ & \phn $  9.777 \pm 0.048  $  &\phn $  7.058 \pm 0.076  $  &\phn $   47.73 \pm   2.33  $  & $    245.7 \pm    29.7  $  &\\
Swift $v$  &   0.55 & \phn 102    &  $ 3.74 \times 10^{ -9}$ & \phn $  9.226 \pm 0.046  $  &\phn $  4.406 \pm 0.071  $  &\phn $   26.42 \pm   2.01  $  & $    154.0 \pm    26.8  $  &\\
$Z$  &   0.90 & \phn\phn 17     &  \nodata  & \phn $  8.834 \pm 0.043  $  &\phn $  2.826 \pm 0.037  $  &\phn $   17.41 \pm   0.72  $  & $    134.7 \pm     5.0  $  &\\
$Y$  &   1.20 & \phn\phn 12     &  \nodata  & \phn $  8.967 \pm 0.043  $  &\phn $  2.673 \pm 0.045  $  &\phn $   15.14 \pm   1.17  $  & $    135.6 \pm    11.1  $  &\\
ROTSE  &   0.60 & \phn 214    &  \nodata  & \phn $  8.866 \pm 0.044  $  &\phn $  3.920 \pm 0.057  $  &\phn $   23.26 \pm   1.00  $  & $     16.6 \pm    11.4  $  &
\enddata
\tablecomments{For each filter, we give the effective wavelength $\lambda$, the number of photometric observations $N$, the flux zero point $\mathcal{F}_{0,i}$, and the global fit parameters $\overline{M}_i$ and $\beta_{n,i}$. Effective wavelengths and flux zero points are taken from the Asiago Database of Photometric Systems \citep{moro00} and \citet{poole08}.}
\label{tab:global}
\end{deluxetable*}

\begin{figure*}
\plotone{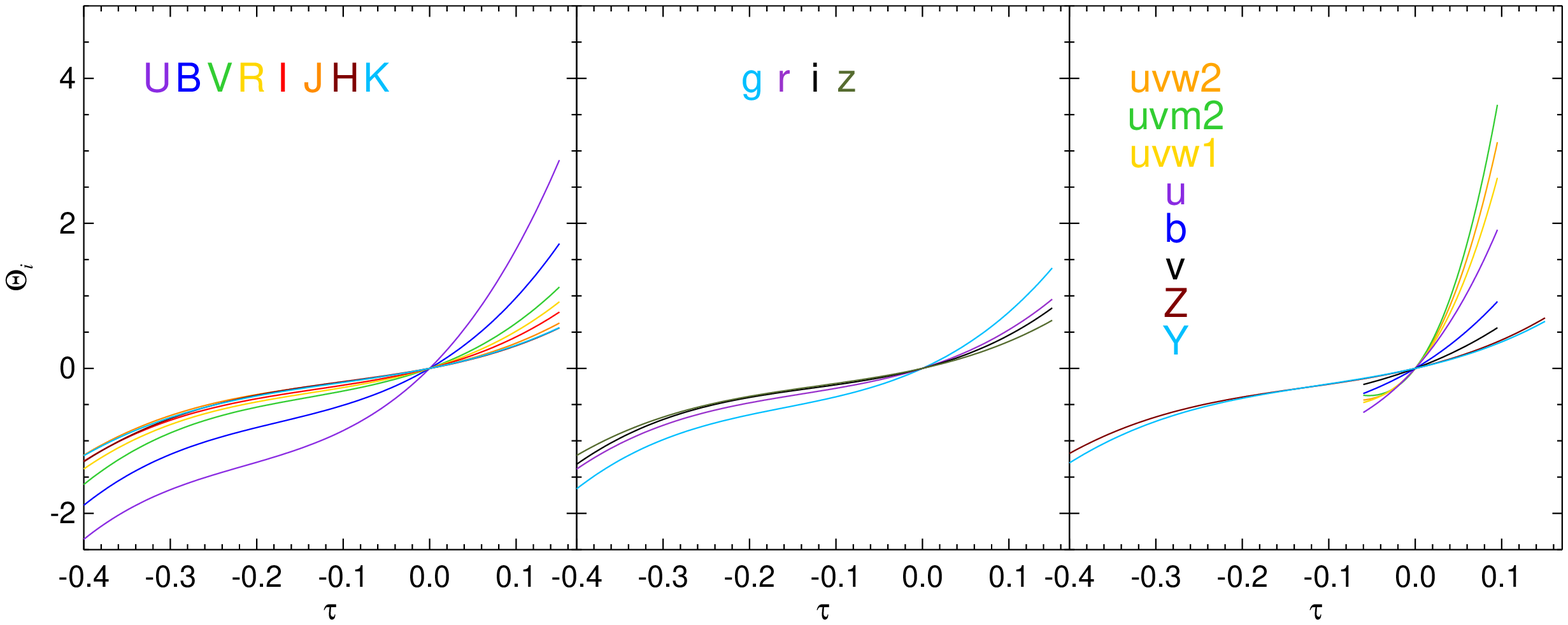}
\caption{The temperature factors $\Theta_i$ as a function of $\tau$. The coefficients are very well ordered in the sense that bluer bands have steeper temperature dependences. The \swift\ bands have data only for $-0.06 \le \tau \le 0.095$ and we thus restrict their plotting range to prevent extrapolation. \label{fig:koef}}
\end{figure*}

In Table~\ref{tab:global}, we present values of the global parameters of the model. We see that the parameters are very well constrained by the data, with typical uncertainties of a few per cent. The only exception is $\beta_{3}$ (Eq.~[\ref{eq:theta}]) for the ROTSE photometric band, where the value is relatively small when compared to $V$ or $R$ band with similar central wavelength and it is comparable to its uncertainty. The reason is that we have ROTSE photometry only for SN2006bp and there is no photometry in other bands during and after the transition, which means that high-order temperature term is poorly constrained in relation to other bands.

In Figure~\ref{fig:koef} we investigate the behavior of the temperature polynomial $\Theta_i$. We see that there is a clear trend with the central wavelength of the filter in the sense that bluer filters have steeper $\Theta_i(\tau)$. The only exception is $K$ band, which is steeper than $J$ and $H$ for $\tau>0$ and the \swift\ $uvw2$ and $uvm2$ bands, which are only shown where data exist to constrain their values. This reflects the time evolution of the spectral energy distribution of Type II-P supernovae, potentially indicating evolution of strong emission or absorption lines. We apply our results on the global parameters to determine supernova spectral energy distributions (Sec.~\ref{sec:sed}), bolometric light curves (Sec.~\ref{sec:bolo}), bolometric corrections (Sec.~\ref{sec:bc}), and compare to theoretical supernova spectrophotometric models through the dilution factors (Sec.~\ref{sec:dilution}).

\begin{figure*}
\centering
\includegraphics[width=\textwidth]{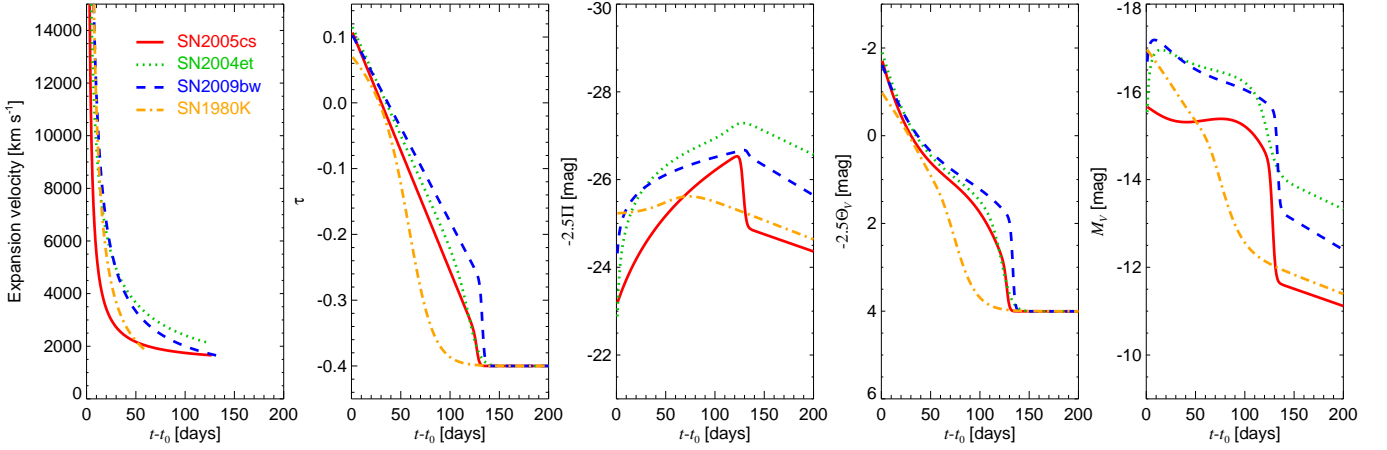}
\caption{Decomposition of the supernova light curves into the photospheric radius and temperature changes. From left to right, the panels show the expansion velocities, temperature parameter $\tau$, photospheric radius function $-2.5\Pi$, and the $V$-band temperature function $-2.5\Theta_V$. Combining the radius and temperature functions with appropriate zero-points gives the absolute $V$-band light curves in the right-most panel. The chosen supernovae illustrate the morphological variations allowed by our model: the flat double-peaked plateau of SN2005cs (solid red),the faster plateau decline of SN2004et (dotted green) and SN2009bw (dashed blue), and the nearly linear light curve of Type II-Linear SN1980K (dash-dotted orange).}
\label{fig:platlin}
\end{figure*}

Now we investigate whether the components of the model have the intended meaning. In Figure~\ref{fig:platlin}, we show the time evolution of the components of Equation~(\ref{eq:main}), specifically the expansion velocity $v(t)$, temperature parameter $\tau$, and the achromatic and chromatic parts of the light curves, $-2.5\Pi$ and $-2.5\Theta_i$. As expected, $v$ and $\tau$ decrease with time in a similar manner in all supernovae. The achromatic component of the light curve $-2.5\Pi$ proportional to the radius increases in time because $\omega_1 > -1$ until $\tp$, when the supernova becomes transparent. After time $\tp$,  $-2.5\Pi$ drops, and the subsequent evolution is dominated by the radioactive decay. The chromatic component of the light curve, $-2.5\Theta_V$, which we evaluate in the $V$ band for the purposes of Figure~\ref{fig:platlin}, decreases similarly to $\tau$, but with additional ``wiggles'' to account for color changes during the plateau. The evolution of $\tau$ and $-2.5\Pi$ is qualitatively similar to the evolution of the color temperature and apparent angular radius presented by \citet[][Fig.~6]{hamuy01}. One would not expect a detailed quantitative match, especially in case of $\tau$, because it is not identical to either the color or effective temperature. This indicates that the individual components of our model indeed have the intended meaning presented in Section~\ref{sec:model} and that they closely match the results obtained in the traditional expanding photosphere method. We discuss the relation between $\tau$ and effective temperature in Section~\ref{sec:bc}.

\begin{figure}
\plotone{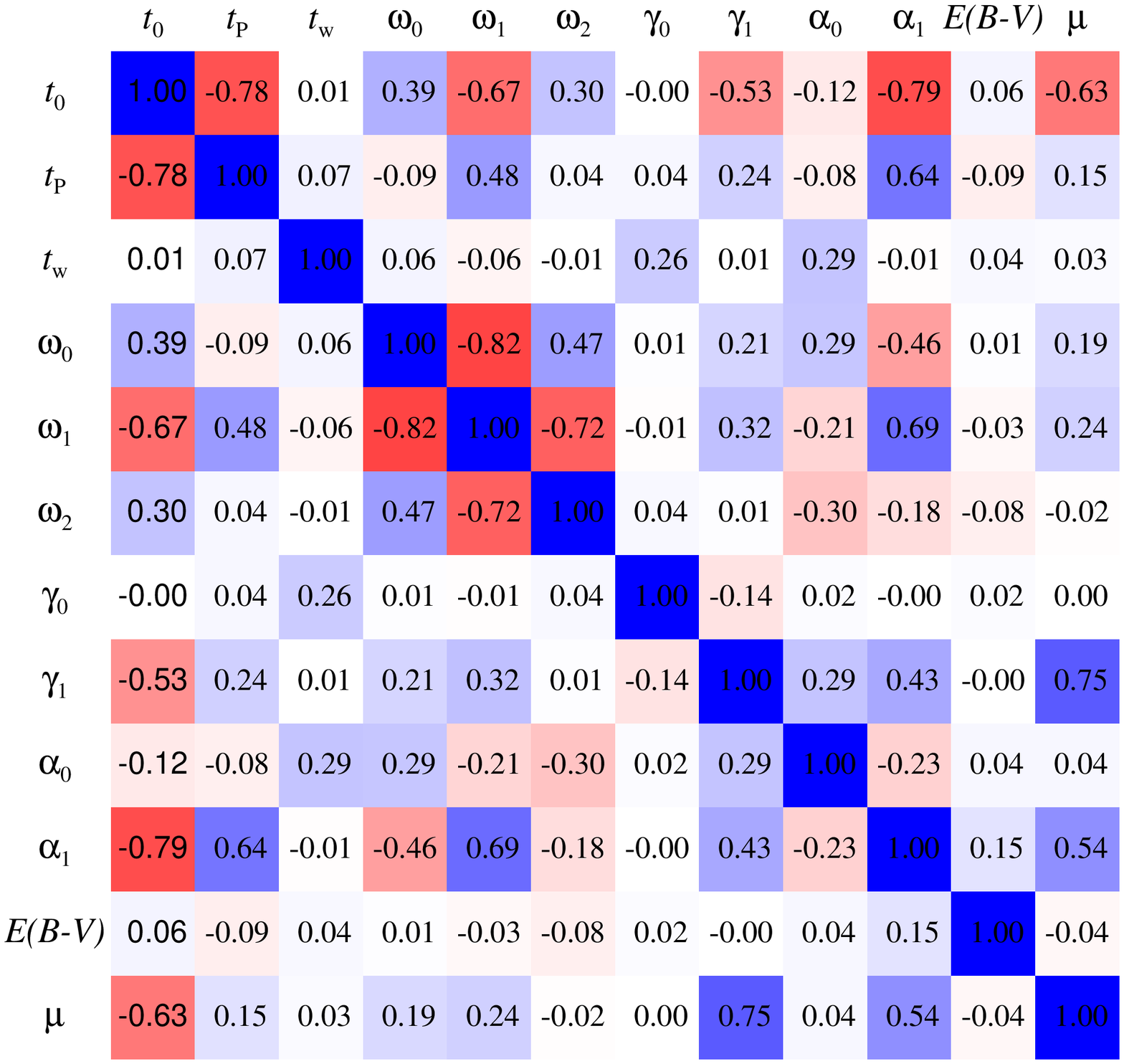}
\caption{Median correlation matrix of the individual supernova parameters. Correlation coefficients are color-coded from $1$ (blue) through $0$ (white) to $-1$  (red), and are also explicitly given by numbers. }
\label{fig:covar}
\end{figure}

With the physics of the model verified, we now ask whether the parameters describing the data are robustly obtained by the fitting process. In Figure~\ref{fig:covar}, we show the correlation matrix of the supernova-specific parameters obtained by calculating the median of each element over all supernovae in our sample. Using the median should mitigate the effect of supernovae lacking some of the observations, which leads to excessively high correlation between some parameters unrelated to the underlying model. For example, it is impossible to constrain all of $\omega_0$, $\omega_1$, and $\omega_2$ for a supernova with less than three velocity measurements. From Figure~\ref{fig:covar} we see that the time of explosion $\texp$ and the plateau duration $\tp$ are anticorrelated mutually and correlated with $\alpha_1$. This can be understood by realizing that supernovae generally lack coverage early on and thus uncertainty in $\texp$ implies uncertainty in the duration of the plateau and the temperature at the moment of explosion $\alpha_1$. The transition width $\tw$ is little correlated with other parameters, as was found also by \citet{anderson14a}. The velocity parameters $\omega_0$, $\omega_1$, and $\omega_2$ are mutually relatively highly correlated. The distance modulus $\mu$ is highly correlated with the explosion time $\texp$ and $\omega_1$. This is similar to the usual expanding photosphere method, where the distance and explosion time are obtained from fitting the apparent angular radius versus expansion velocity: uncertainty in the distance directly translates to uncertainty in the explosion time. These correlations imply that to precisely determine the distance of a Type II-P supernova, we require photometric observations constraining the explosion time and velocity measurements spanning long-enough interval so that the exponent $\omega_1$ is well-constrained. The highest correlation of $E(B-V)$ with other parameters is $0.07$, which indicates that we robustly determine the reddening with little influence from other supernova-specific parameters.

Our method works so well for determination of $E(B-V)$ not only because we have multi-band photometry for many of our objects. If a supernova has enough velocity measurements or the distance is well-known, the radius $\Pi$ is constrained and changing $\tau$ results not only in changes of the color, but also in changes of the supernova magnitude. In other works, making the supernova redder to mimick the reddening would also make it fainter, with constant radius. Keeping the global parameters fixed, we fitted the individual supernova parameters of SN2004et and SN2009N based on a dataset that includes only velocity and $B$ and $V$ measurements. We found that in this case $E(B-V)$ was $\sim 20\%$ smaller than what we obtain in Table~\ref{tab:sn_pars}. This opens a path to reliable determination of supernova reddenings. Note that our determination of $E(B-V)$ is potentially biased, because the model will try to absorb differences between supernovae not parameterized in the model (such as the metallicity) into changes of $E(B-V)$ and other parameters, as we have shown for Cepheids in \citet{pejcha12}.

\begin{figure}
 \plotone{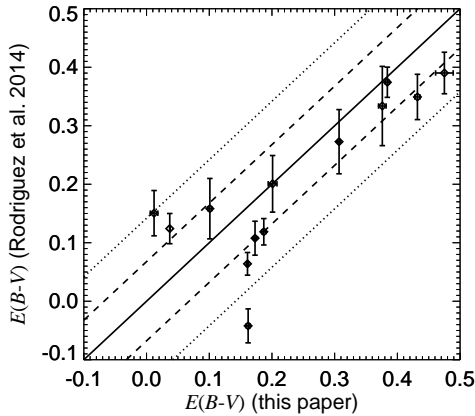}
\caption{ Comparison of our estimates of $E(B-V)$ with the results of \citet{rodriguez14}, where we sum their Milky Way reddenings with the results of their C3($BVI$) method for host galaxy extinction assuming $\rv=3.1$ reddening law. The solid line marks a perfect correlation between the results. We show $1\sigma$ confidence regions of perfect correlation between the C3($BVI$) method and spectrum-fitting estimates of \citet{rodriguez14} for their full sample (dotted lines) and with outliers removed (dashed lines). }
\label{fig:comp_ebv}
\end{figure}

In Figure~\ref{fig:comp_ebv}, we compare our estimates of $E(B-V)$ to results of \citet{rodriguez14}, which are based on color standardization of Type II supernovae using their C3($BVI$) method. The overlap of our sample and the sample of \citet{rodriguez14} is 13 objects and we see relatively good agreement, especially at high $E(B-V)$. At lower reddenings, the discrepancies are greater. We also show $1\sigma$ regions of perfect correlation between the C3($BVI$) results and spectrum-fitting estimates determined by \citet{rodriguez14} for all of their supernovae and with outliers removed. All of the supernovae fall within these regions, except one supernova, where \citet{rodriguez14} obtained significantly negative $E(B-V)$. This suggests that our reddening estimates are definitely not worse than previous or contemporary results. We emphasize that our results make use of all available colors, while \citet{rodriguez14} use only $BVI$ pass bands.

\subsection{Morphology of light curves and the differences between plateau and linear supernovae}
\label{sec:iil}

Recently, attention has been devoted to understand the morphology of Type II supernova light curves with properties ranging from flat plateaus to steep declines \citep{anderson14a,faran14,sanders14}. In particular, \citet{arcavi12} suggested the possibility that there are no supernovae with properties in between Type II-P and Type II-Linear supernovae. Conversely, \citet{anderson14a} and \citet{sanders14} found a continuum of light curves of Type II supernovae. Since our model can disentangle the observed data into radius and temperature variations, we decided to investigate these issues within our model.

\begin{figure}
\plotone{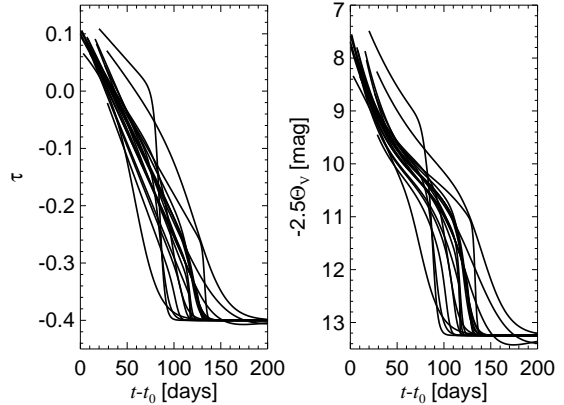}
\caption{Temperature parameter $\tau$ (left panel) and the chromatic part of the light curve in $V$ band $-2.5\Theta_V$ (right panel) as a function of time after explosion. We show only well-observed supernovae with distance modulus uncertainties smaller than $0.2$\,mag and the curves start at the time of the first observation of each supernova.}
\label{fig:temp}
\end{figure}

In Figure~\ref{fig:platlin} we show three Type II-P supernovae with the different shapes of the plateau: the relatively steeply declining SN2009bw, the flat plateau of SN2005cs with a noticeable bump just before the transition, and the intermediate case of SN2004et. We see that the overall behavior of velocity and chromatic part of the light curve $-2.5\Theta_V$ are very similar for these supernovae. In fact, this is true for all supernovae with enough data, as we show in Figure~\ref{fig:temp}.  The small differences in the velocity evolution, however, translate into different steepness for the achromatic part of the light curve $-2.5\Pi$. In SN2004et and SN2009bw, there is initially a fast rise in the radius, which then gradually slows down. As a result, the decrease of brightness due to $-2.5\Theta_V$ dominates and we observe a declining plateau. In SN2005cs, the radius is increasing fast\footnote{Note that the expansion velocity and hence radius of SN2005cs is consistently smaller than those of SN2004et and SN2009bw.} even before the transition, which almost exactly compensates for the decrease of brightness due to the decreasing temperature. As a result, we observe a flat plateau with a bump. Since the temperature evolution is so similar, the morphological differences in the Type II-P supernova light curves are thus due to different photospheric velocity evolution.

\begin{figure}
\plotone{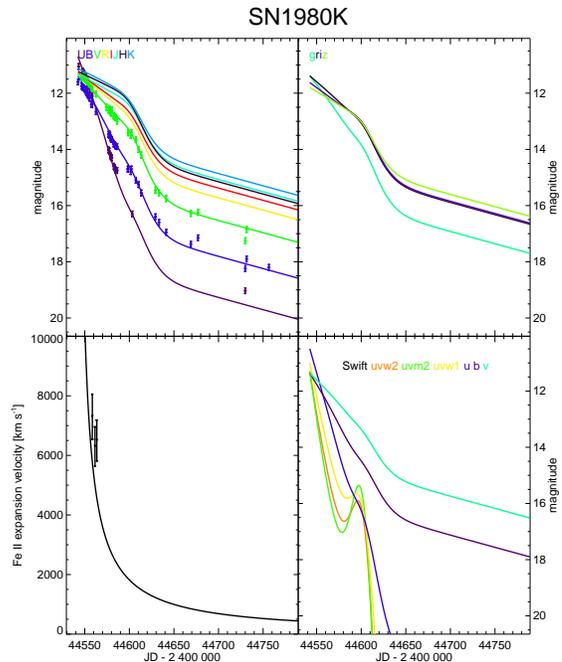}
\caption{Light curve and expansion velocity fits for Type II-Linear SN1980K.}
\label{fig:sn1980k}
\end{figure}

A further understanding of the light curve morphology can be obtained by applying our model to Type II-Linear supernovae. The only Type II-Linear supernova we found in the literature with enough data for a successful fit is SN1980K \citep{barbon82,buta82}. SN1980K is convenient for our purposes, because it exploded in NGC6946 with a distance very well constrained by SN2004et (Fig.~\ref{fig:dist_ned} and Table~\ref{tab:galaxy}) and we do not have to worry about uncertainties in the overall radius scaling. In Figure~\ref{fig:sn1980k} we show the light curve and expansion velocity fits, and Figure~\ref{fig:platlin} shows the decomposed changes in radius and temperature. We obtained a reasonable fit to the data. The estimated optically-thick phase duration of $\tp \approx 61$\,d is noticeably shorter than for normal plateau supernovae. However, the transition width is fairly long, $\tw \approx 13$\,d, and as a result the temperature parameter reaches values typical for exponential decay ($\tau = \tau_{\rm thick} \equiv -0.4$) about $100$\,days after the explosion, which is more similar to ordinary Type II-P supernovae. Still, after $\texp+50$\,d, the temperature falls faster than in normal plateau supernovae. This is consistent with the finding of \citet{anderson14a} that faster declining supernova have shorter duration of the optically-thick phase. The radius of SN1980K stays almost constant during the optically-thick phase, but this is mostly because of the large $\tw$, which makes the supernova nearly transparent for most of the observed time. As a result, the steeply declining light curve is mainly due to temperature changes of the photosphere. The successful fit and the resulting parameters indicate that Type II-Linear supernovae can be modelled along with normal Type II-P explosions and that Type II-Linear brightness variations are driven primarily by changes in the photospheric temperature.

\begin{figure}
\plotone{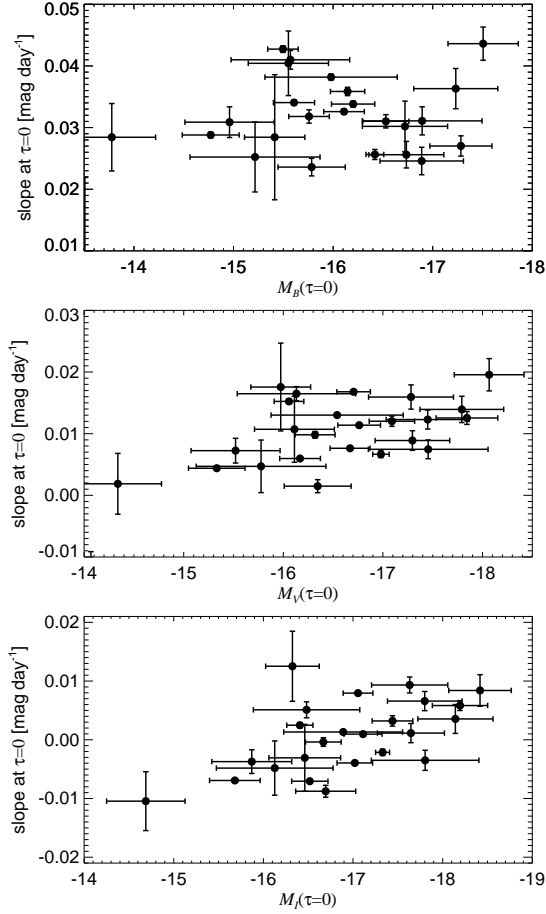}
\caption{Slope of the light curve (Eq.~[\ref{eq:slope}]) during plateau at $\tau=0$ in the $B$ (top), $V$ (middle), and $I$ (bottom) bands as a function of absolute magnitude. The $V$ and $I$ bands show the correlation reported by \citet{anderson14a}, but the $B$ band does not. Uncertainties in the parameters come from the modified fit with $\mathcal{H}/\ndof = 1$.}
\label{fig:ms2}
\end{figure}

To put the discussion of the light curve morphology on more quantitative grounds, we address the recent finding of \citet{anderson14a} that the $V$-band maximum magnitude and the plateau rate of magnitude decline are correlated. The maximum magnitude in our model is not easy to obtain analytically, but we instead consider the absolute magnitude at $\tau=0$, $M_i(\tau=0) = \overline{M}_i - 2.5\Pi$. The epoch corresponding to $\tau=0$ is $t_{\tau=0}-\texp =-\alpha_1/\alpha_0$ (Eq.~[\ref{eq:tauthick}]) and the $BV$ color temperature at this point is $\sim 6400$\,K using the relations between color index and temperature of \citet{hamuy01} as implemented below in Section~\ref{sec:dilution}. The effective temperature at $\tau=0$ is $\sim 5600$\,K as discussed in Section~\ref{sec:bc}. The light curve slope at $\tau=0$ is
\beq
\left.\frac{\intd M_i}{\intd t}\right|_{\tau=0} \approx -\frac{5}{\ln 10}\left(\frac{1}{v}\frac{\intd v}{\intd t}- \frac{\alpha_0}{\alpha_1}  \right) - 2.5\beta_{1,i}\alpha_0,
\label{eq:slope}
\eeq
where the approximation comes from the assumption of $w=0$.

We show the absolute magnitude as a function of light curve slope in Figure~\ref{fig:ms2} for $B$, $V$, and $I$ bands. The uncertainties in both quantities were obtained by applying Equation~(\ref{eq:uncert_prop}) to Equation~(\ref{eq:slope}) with the full covariance matrix of the fit. We see that we reproduce the $V$ band correlation of \citet{anderson14a} and that this correlation exists also for the $I$ band, but not for the $B$ band. In general, we find that the correlation does not exist for bands with $\lambda \lesssim 0.5\,\mu$m, in other words, the correlation exists only when the first term in Equation~(\ref{eq:slope}) dominates, which occurs when $\beta_{1,i}$ is small. Why does this correlation exist? Early after explosion, $\omega_2$ in Equation~(\ref{eq:v}) can be neglected, which gives $M_i(\tau_0) \sim -5\log\omega_0 - 5(\omega_1+1)\log(-\alpha_1/\alpha_0)$, $v^{-1}\intd v/\intd t \sim (\omega_1+1)/t$, and $\intd M_i /\intd t \sim (\alpha_0/\alpha_1)(\omega_1+1)$. In other words, both quantities are roughly proportional to $\omega_1+1$ leading to a correlation that is not perfect because of the other terms in both expressions.

Clearly, the parameter $\omega_1$, measuring the curvature of the velocity decrease, plays an important role in determining the morphology of Type II-P supernovae. To gain insight about the physical property of the exploding medium controlling $\omega_1$, we consider an homologously expanding adiabatic medium with density profile $\rho \propto (r/R_{\rm out})^{-k} R_{\rm out}^{-3}$, where the outer boundary $R_{\rm out} \propto t$ is expanding linearly in time. We assume that the medium is dominated by radiation and that the temperature is constant as a function of radius giving $T \propto t^{-1}$ \citep{arnett80}. Assuming a general opacity law $\kappa \propto \rho^nT^{-s}$, the photospheric radius $R$ in such medium is determined by
\beq
\tau =\int_{R}^{R_{\rm out}} \kappa\rho\intd r =  \frac{2}{3}.
\eeq
Using our expressions for density, opacity, and temperature, and assuming that $R_{\rm out} \gg R$, we obtain the rate of photospheric expansion
\beq
\frac{\intd \ln R}{\intd \ln t} = \frac{(k-3)(n+1)+s}{k(n+1)-1}.
\label{eq:dlnr_theo}
\eeq
Equations~(\ref{eq:homo}--\ref{eq:v}) give the rate of photospheric expansion from observed quantities as
\beq
\frac{\intd \ln R}{\intd \ln t} = 1+ \frac{\omega_0\omega_1 t^{1+\omega_1}}{\omega_0t^{1+\omega_1} + \omega_2 t},
\label{eq:dlnr_obs}
\eeq
which reduces to $\intd \ln R/\intd \ln t = 1+\omega_1$ if $\omega_2$ is negligible. The opacity law can be either Thompson opacity ($n=s=0$), Kramers opacity ($n=1$, $s=3.5$) or H$^-$ opacity ($n=0.5$, $s=-7.7$), where the latter is perhaps most appropriate at $\tau=0$ with effective temperature of about $5600$\,K. The density exponent implied by Equations~(\ref{eq:dlnr_theo}--\ref{eq:dlnr_obs}) for H$^-$ opacity is $k \approx 15.3$, $13.1$, and $10.4$ for SN2005cs, SN2004et, and SN2009bw, respectively, at $t+\texp=30$\,days. This is in relatively good agreement with the density exponents in the theoretical models of \citet{dh08,dh11}. As can be seen in Figure~\ref{fig:platlin}, steeper density profiles correspond to flatter plateaus.

Since the temperature evolution and thus the chromatic part of the light curve are very similar for all supernovae (Fig.~\ref{fig:temp}), this implies that {\em for a fixed opacity law, the shape of the supernova plateau is controlled by the homologous density profile of the ejecta, with steeper density profiles resulting in flatter light curves}.  The supernova ejecta have also temperature structure that evolves in time so that further insight into this issue requires much more realistic models than the simplistic estimate presented here. The relation of the light curve morphology to the mass of hydrogen has also been proposed by \citet{anderson14a,anderson14b}. Nickel mixing in the ejecta can have also effect on the morphology of the optically-thick phase \citep{bersten11,kasen09}.

\subsection{Distances}
\label{sec:dist}

\begin{deluxetable}{cccc}
\tabletypesize{\footnotesize}
\tablecolumns{2}
\tablewidth{0pc}
\tablecaption{Galaxy Distances}
\tablehead{ \colhead{Galaxy} & \colhead{$\mu$ [mag]} &\colhead{$\sigma_\mu$} & \colhead{$\sigma'_\mu$\tablenotemark{a}}}
\startdata
M51~(NGC5194)         &  $ 29.460 $ & $ 0.112 $ & $0.282$ \\
M61~(NGC4303)         &  $ 31.363 $ & $ 0.646 $ & $0.663$ \\
M95~(NGC3351)         &  $ \equiv 30.000 $ & & \\
MCG-01-04-039         &  $ 35.679 $ & $ 0.140 $ & $0.433$ \\
MCG-01-32-035         &  $ 34.905 $ & $ 0.220 $ & $0.449$ \\
NGC0918               &  $ 33.275 $ & $ 0.279 $ & $0.621$ \\
NGC1637               &  $ 30.088 $ & $ 0.057 $ & $0.196$ \\
NGC2139               &  $ 32.105 $ & $ 0.195 $ & $0.478$ \\
NGC2403               &  $ 27.895 $ & $ 0.136 $ & $0.373$ \\
NGC3184               &  $ 29.929 $ & $ 0.057 $ & $0.195$ \\
NGC3239               &  $ 30.018 $ & $ 0.044 $ & $0.148$ \\
NGC3389               &  $ 32.005 $ & $ 0.143 $ & $0.448$ \\
NGC3953               &  $ 31.085 $ & $ 0.084 $ & $0.229$ \\
NGC4027               &  $ 31.458 $ & $ 0.117 $ & $0.389$ \\
NGC4088               &  $ 30.184 $ & $ 0.300 $ & $0.595$ \\
NGC4487               &  $ 31.517 $ & $ 0.058 $ & $0.204$ \\
NGC4651               &  $ 30.436 $ & $ 0.239 $ & $0.586$ \\
NGC5377               &  $ 33.119 $ & $ 0.143 $ & $0.418$ \\
NGC5777               &  $ 31.895 $ & $ 0.186 $ & $0.461$ \\
NGC6207               &  $ 31.460 $ & $ 0.096 $ & $0.335$ \\
NGC6946               &  $ 28.265 $ & $ 0.068 $ & $0.210$ \\
NGC7793               &  $ 28.385 $ & $ 0.504 $ & $0.699$ \\
UGC02890              &  $ 30.777 $ & $ 0.048 $ & $0.165$ \\
UGC12846              &  $ 32.072 $ & $ 0.096 $ & $0.307$ \\

\enddata
\tablenotetext{a}{Uncertainty $\sigma'_\mu$ obtained from a fit with the uncertainties of the data rescaled to give $\mathcal{H}/\ndof = 1$.}
\tablecomments{See text for comments on individual galaxies.}
\label{tab:galaxy}
\end{deluxetable}

\begin{figure*}
\centering
\includegraphics[width=\textwidth]{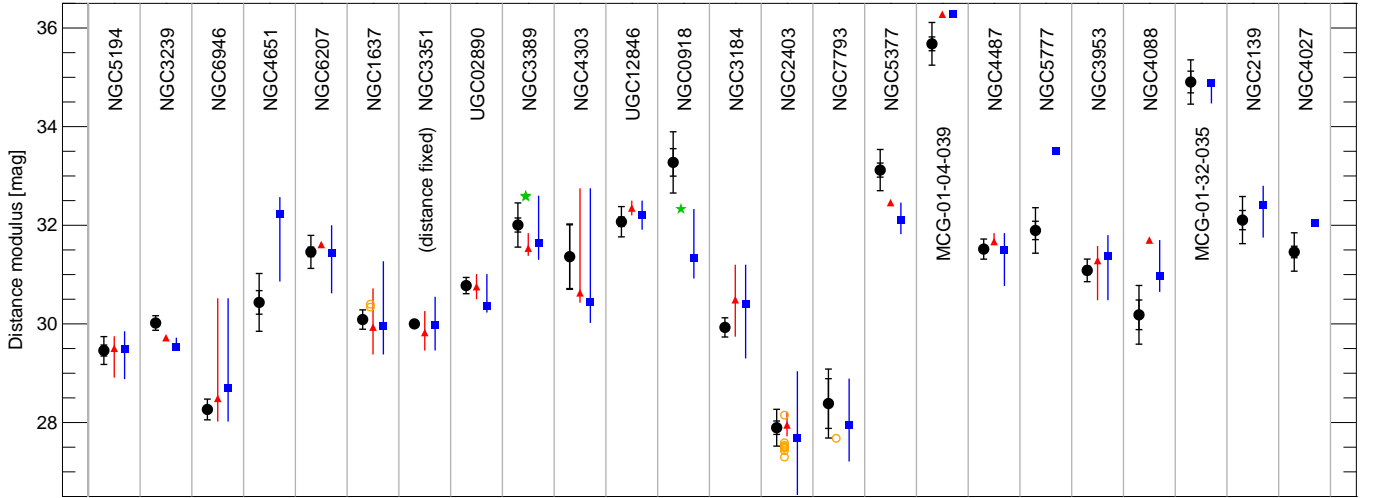}
\caption{Comparison of the distance estimates of galaxies in our sample (solid black circles with $1\sigma$ uncertainties of the original fit and the modified fit with $\mathcal{H}/\ndof = 1$) to the data in NED: median values of either Type II supernovae only or all distance estimates are shown with solid red triangle or blue square, respectively, while the full range of measurements is shown with vertical colored lines. For a few galaxies, we highlight individual distance measurements with Cepheids (open orange circles) or with supernovae Ia (solid green stars). Distances of Type II supernovae are from \citet{baron96,baron07,bartel88,bose14,dessart06,dessart08,elhamdi03,fraser11,hamuy01,hendry06,inserra12b,iwamoto94,jones09,leonard02a,leonard02b,leonard03,olivares10,poznanski09,richmond96,roy11,sahu06,schmidt92,schmidt94a,schmidt94b,sparks94,takats14,takats06,takats12,tomasella13,vinko06,vinko12,weiler98}, of Cepheids are from \citet{freedman88,freedman01,humphreys86,leonard03,madore91,mcalary84,metcalfe91,pietrzynski10,saha06}, and of Type Ia supernovae from: \citet{maguire12,parodi00}.}
\label{fig:dist_ned}
\end{figure*}

In Table~\ref{tab:galaxy}, we present the distance estimates to individual galaxies in our sample. We compare our results to the entries in the NASA Extragalactic Database (NED), focusing on the distance measurements from Type II supernovae, Cepheids, and Type Ia supernovae in Figure~\ref{fig:dist_ned}. Overall, we see relatively good agreement in the sense that our results fall within the range of previous distance determinations. The overall distance scale can be moved by changing the distance to NGC3351 (host of SN2012aw), which we assume to be $\mu \equiv 30.00$\,mag. We point out that in principle, our method allows determinations of distances to supernovae with no velocity measurement by applying priors on $\omega_0$, $\omega_1$, and $\omega_2$. However, in this case the uncertainties are rather large ($\gtrsim 0.5$\,mag) and there is a great potential for systematic offsets.

We now specifically discuss several galaxies where there is substantial disagreement with previous results. The photometric and spectroscopic observations of SN2006my in NGC3239 start shortly before the transition to optically-thin phase. As a result, $\texp$ is not well constrained and thus $\mu$ is potentially substantially biased as we discussed in Figure~\ref{fig:covar}. This particular aspect can be improved by applying constraints on the age of the supernova from the appearance of the spectrum while treating appropriately the uncertainty associated with this prior information. Including such constraints is beyond the scope of this paper.

Our distance for NGC3389 (host of SN2009md) is noticeably smaller than the one measured from the Type Ia supernova SN1967C \citep{parodi00}, but agrees reasonably well with previous estimates from SN2009md \citep{fraser11,bose14}. SN2009md has good spectroscopic and multi-band photometric coverage so the disagreement comes perhaps from the inaccurate photographic photometry of SN1967C. Our distance for NGC918 (host of SN2009js) is significantly larger than that based on the recent Type Ia supernova SN2011ek in the same galaxy \citep{maguire12}. SN2009js is relatively well-observed in $BVRI$, but has only one velocity measurement so that the distance estimate is more susceptible to systematic errors. The agreement is almost within the uncertainty, if the rescaled fit with $\mathcal{H}/\ndof = 1$ is used. There are no previous distance estimates to SN2009js in NED. Our distance to NGC7793 is based on SN2008bk, for which we do not have any velocity measurements and only $V$-band photometry during the plateau and thus the relatively good agreement with NED distances is rather surprising. Our distance estimate almost agrees with Cepheid distance of \citet{pietrzynski10}. The distance to NGC5377 is based on SN1992H, which has relatively good velocity measurements, but rather sparse $BVR$ photometry. The distance to MCG-01-04-039 is based on SN1992am with $BVI$ photometry starting about $25$\,days after $\texp$. Our lack of K-corrections probably plays some role, but most of the disagreement can probably be attributed to the poor early coverage. Using the fit with $\mathcal{H}/\ndof=1$ yields higher distance uncertainty and thus better agreement with previous measurements. Photometry of SN2001dc in NGC5777 does not cover the first $\sim 35$\,days and the first velocity measurement is $\sim 50$\,days after $\texp$, which explains the disagreement. SN2009dd in NGC4088 has well-determined $\texp$, but not enough velocity measurements with sufficient precision. The prior supernova distance determination of NGC4088 was with a different Type II SN1991G \citep{poznanski09}.

To summarize, Type II-P supernovae can yield solid distances provided there is multi-band photometry, good expansion velocity coverage and that the explosion time can be well constrained from the observations.

\subsection{Spectral energy distribution}
\label{sec:sed}

\begin{figure*}
\plottwo{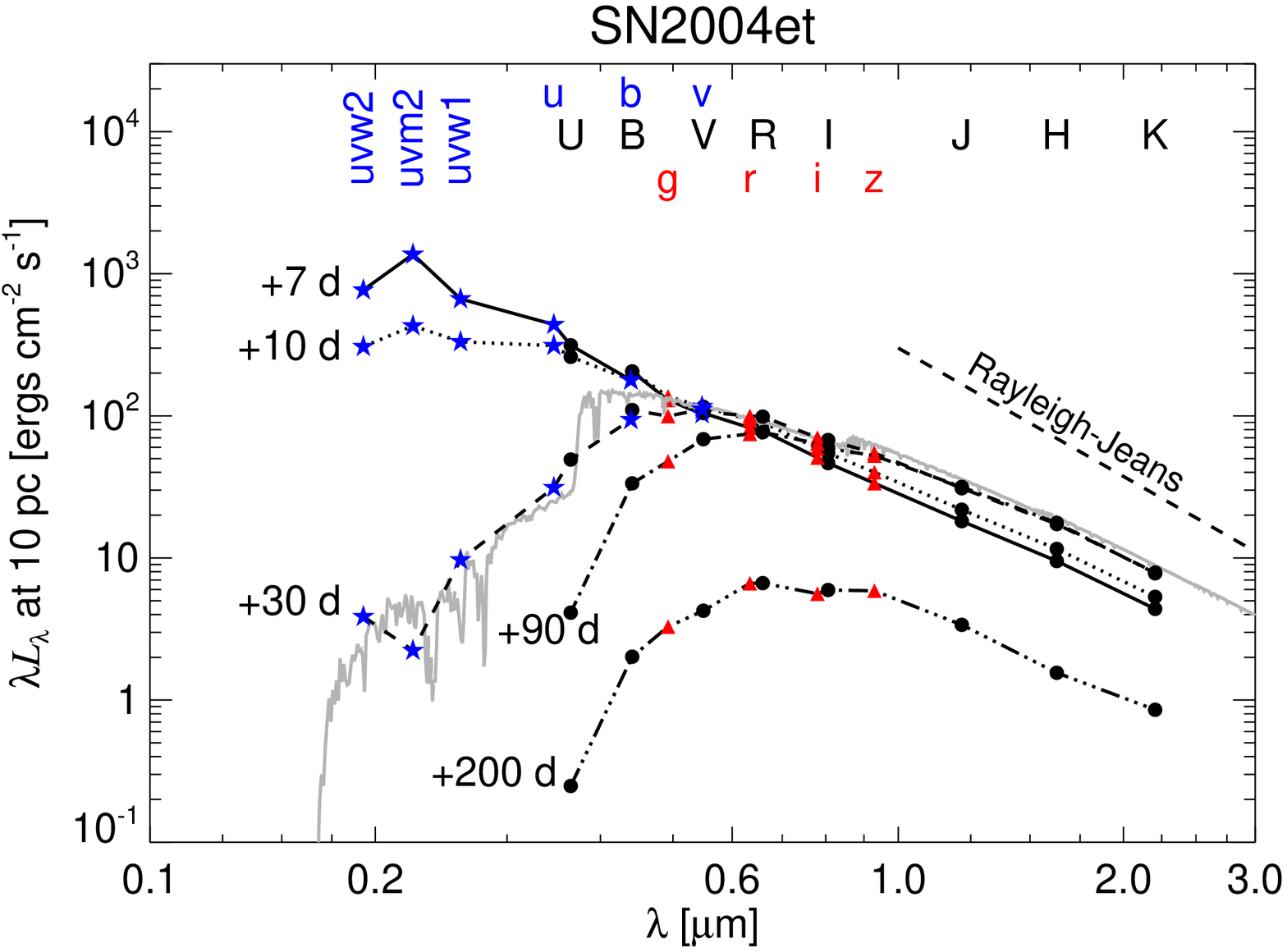}{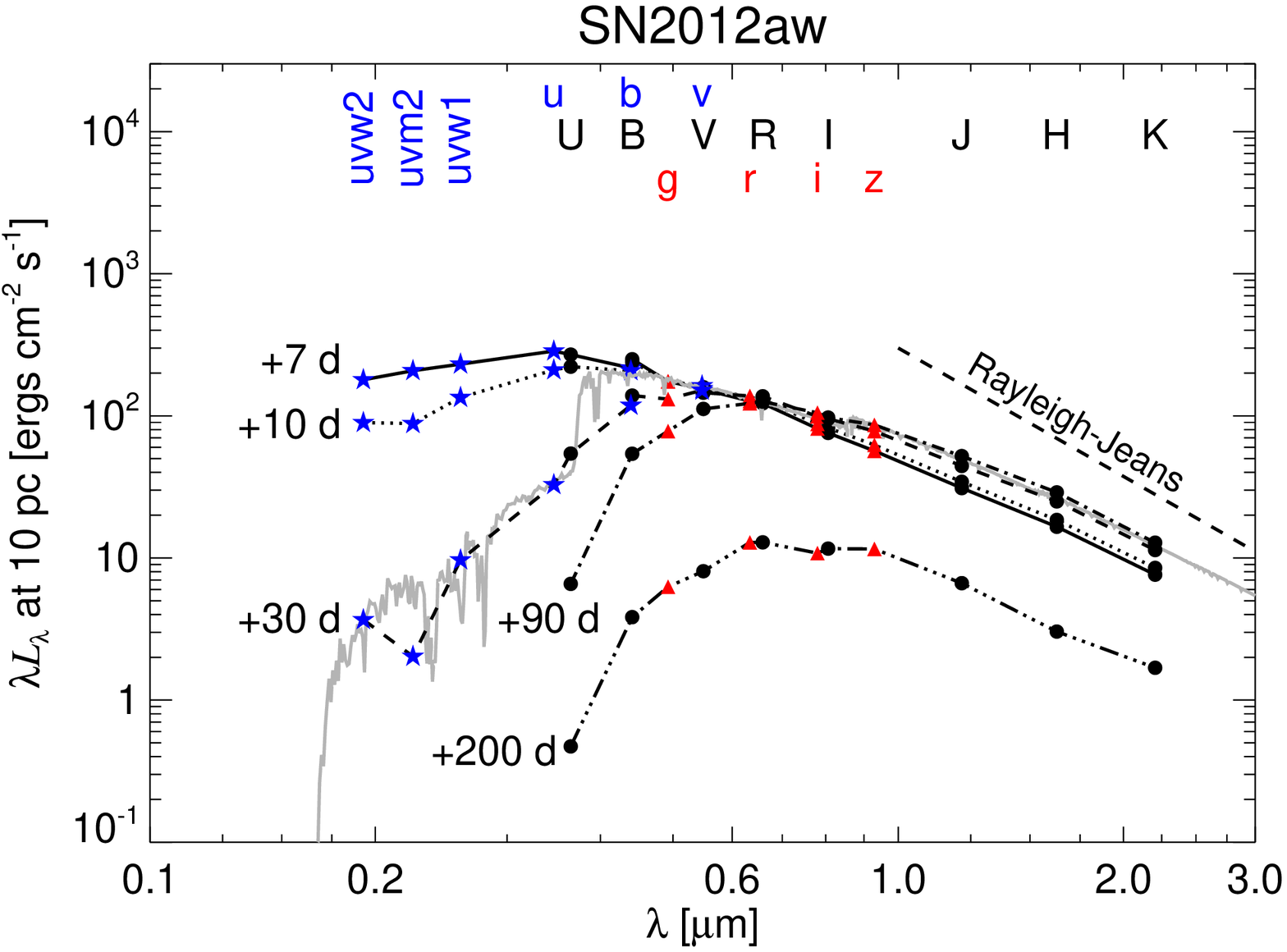}
\caption{Absolute reddening-corrected SEDs of SN2004et (left) and SN2012aw (right) constructed from our model. We show the SEDs at five epochs: $7$ (solid line), $10$ (dotted), $30$ (dashed), $90$ (dash-dotted), and $200$\,days (dash-dot-dot) after the explosion. The SEDs are constructed from the $UBVRIJHK$ (black circles), $griz$ (red triangles) and \swift\ bands (blue stars). The dashed line indicates the black body Rayleigh-Jeans with $\lambda F_\lambda \propto \lambda^{-3}$. The grey line shows theoretical SED of an F0 supergiant with $T_{\rm eff} = 7000$\,K and $\log g = 0.5$ \citep{castelli04} normalized to the $V$-band flux at $\texp+30$\,days. }
\label{fig:sed}
\end{figure*}

Within our model, the monochromatic luminosity $L_i$ at the central wavelength $\lambda_i$ of band $i$ at a distance of 10\,pc ($\mu=0$) is 
\beq
L_i = 10^\Pi \mathcal{F}_i,
\label{eq:fi}
\eeq
where $\mathcal{F}_i$ is a monochromatic flux at a specific radius $\Pi=0$ defined as
\beq
\mathcal{F}_i =  \mathcal{F}_{0,i} 10^{-0.4(\overline{M}_i - 2.5\Theta_i)},
\label{eq:ffi}
\eeq
where the flux zero points $\mathcal{F}_{0,i}$ are taken from the Asiago Database of Photometric Systems \citep{moro00} and from \citet{poole08} for the \swift\ bands and are given in Table~\ref{tab:galaxy}. The usual process of dereddening and correcting for the supernova distance is achieved in our model simply by not including terms $\mu$ and $\ri E(B-V)$ in Equation~(\ref{eq:main}).

In Figure~\ref{fig:sed}, we show absolute spectral energy distributions (SEDs) of SN2004et and SN2012aw at several epochs calculated using Equations~(\ref{eq:fi}--\ref{eq:ffi}). We see that the SEDs get progressively redder as the supernova evolution proceeds. Early in the supernova evolution, the flux is dominated by the near-UV emission. In the first two epochs, the fluxes in the three bluest \swift\ bands are relatively flat and do not show any sign of turning down at shorter wavelengths indicating that the peak of the emission is located at even bluer wavelengths ($T_{\rm eff} \sim  6\times 10^4$\,K). It is uncertain how to extrapolate the flux to shorter wavelengths. At about $\texp+30$\,days, the SEDs of both supernovae have $T_{\rm eff} \sim 5800$\,K and are similar to an F0 supergiant. The near-UV emission is already unimportant at this point. At later epochs, the SED achieves a constant shape ($\tau_{\rm thick} \equiv -0.4$) very different from a normal star and only the overall normalization changes in time ($T_{\rm eff} \sim 2500$\,K). At all epochs, the SEDs at $\lambda \gtrsim 1\,\mu$m are very similar to the Rayleigh-Jeans tail of the black body, as indicated by the $\lambda F_\lambda \propto \lambda^{-3}$ dashed line in Figure~\ref{fig:sed}. The relation between $\tau$ and effective temperature will be discussed in Section~\ref{fig:bc}.

\subsection{Bolometric light curves}
\label{sec:bolo}

The bolometric luminosity $\lbol$ is 
\beq
\lbol = 10^{\Pi} \mathcal{F}_{\rm bol},
\label{eq:lbol}
\eeq
where $\fbol$ is the bolometric flux at radius $\Pi=0$
\beq
\mathcal{F}\bol =  \int \mathcal{F}_i \intd \lambda.
\label{eq:fbol}
\eeq
We perform the integral in Equation~(\ref{eq:fbol}) using the trapezoidal rule assuming monochromatic fluxes $\mathcal{F}_i$ at central wavelengths $\lambda_i$ given in Table~\ref{tab:global}. We correct $\fbol$ for longer wavelength infrared flux by extrapolating the $K$-band flux to $\lambda \rightarrow \infty$ assuming the Rayleigh--Jeans tail of the black body, which corresponds to adding $\mathcal{F}_K \lambda_K/3$ to Equation~(\ref{eq:fbol}). This correction is several per cent during the exponential decay and less than $1\%$ just after the explosion. The UV correction is potentially much more important, especially early after the explosion, as can be seen from Figure~\ref{fig:sed}. However, the SED probably peaks at even shorter wavelengths than the bluest of the \swift\ bands and we do not have any constraining data. As a result, we do not perform any UV correction to $\mathcal{F}\bol$ and simply truncate the integral for $\lambda < 0.19\,\mu$m. We use Equation~(\ref{eq:uncert_prop}) to calculate the uncertainties in $\lbol$, fully accounting for all covariances in the model.

\begin{figure}
\plotone{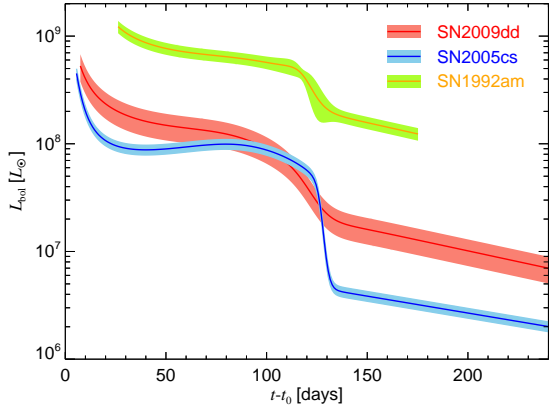}
\caption{Bolometric light curves of SN2009dd (red), SN2005cs (blue), and SN1992am (green/orange) with the accompanying uncertainties calculated with the full covariance matrix of our fit.}
\label{fig:lbol_err}
\end{figure}

In Figure~\ref{fig:lbol_err} we show the evolution of $\lbol$ for three supernovae to illustrate the procedure. We show the bolometric curves starting from the first observation to avoid any extrapolation in our model. At most epochs, the uncertainty is dominated by the uncertainty in the supernova distance. Our prescription, however, can exhibit more complicated behavior, as shown during the transition of SN1992am, which is poorly covered by observations.

\begin{figure*}
\plotone{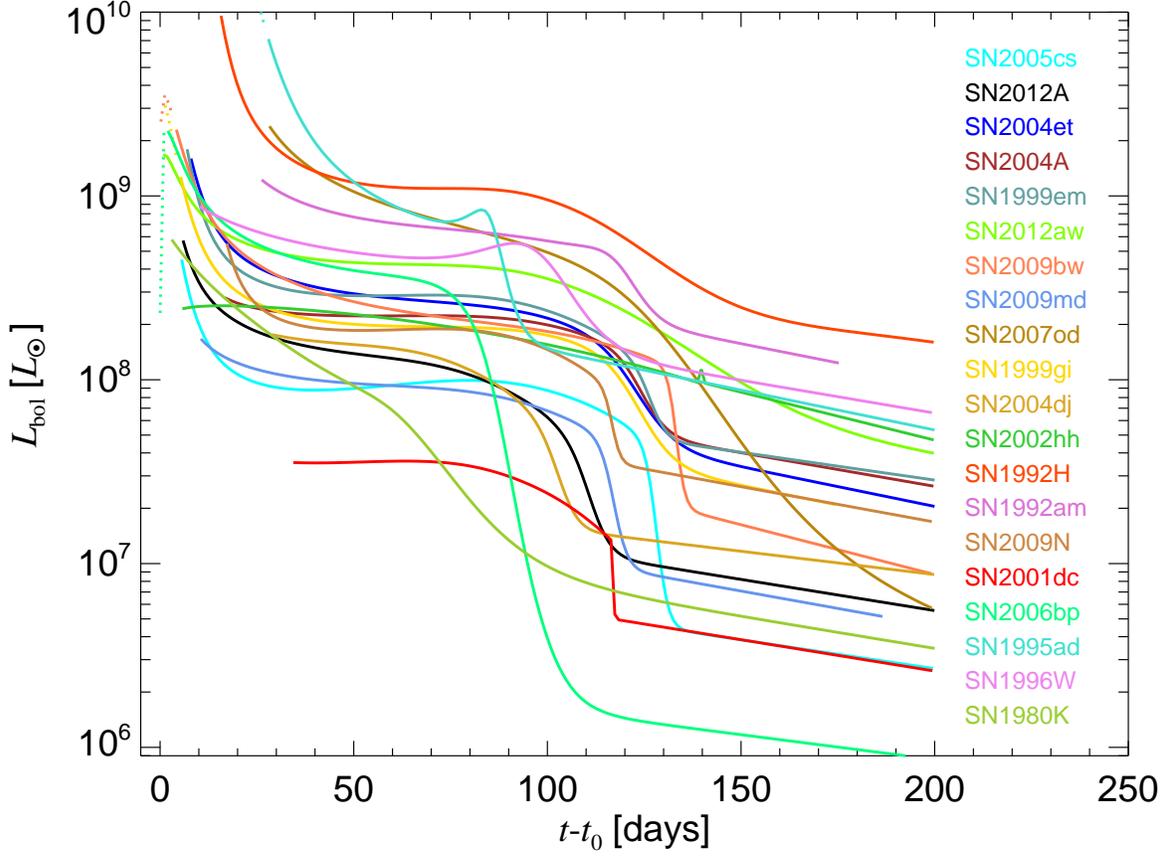}
\caption{Bolometric luminosity as a function of time for supernovae with $\sigma_\mu < 0.2$\,mag. The bolometric luminosity was calculated from the $UBVRIJHK$ bands combined with \swift\ bands $uvw2$, $uvm2$, $uvw1$, and $u$. For each supernova, we show $\lbol$ between the first and last observation. Dotted segments of $\lbol$ mark the temperatures not constrained by \swift\ data ($\tau < 0.095$). The bolometric light curves with uncertainties of individual supernovae are available in the on-line version of Figure~\ref{fig:lc}.}
\label{fig:lbol}
\end{figure*}

In Figure~\ref{fig:lbol}, we show the bolometric light curves of all our supernovae with $\sigma_\mu \le 0.2$\,mag. For the sake of clarity, we do not display the associated uncertainties. We see that we are able to reproduce the previous results on the supernova luminosities. Specifically, we identify SN2001dc, SN2005cs, and SN2009md as low-luminosity supernovae \citep{pastorello04,pastorello06,pastorello09,fraser11}, and SN1992am and SN1996W as relatively luminous objects \citep{schmidt94a,inserra13}. Our model also indicates that SN1992H is luminous, contrary to \citet{clocchiatti96} who found rather normal luminosity. The difference comes from the fact that we obtain a significantly earlier explosion time and thus a larger distance resulting in higher supernova luminosity. We also mention the peculiar case of SN2006bp, which has a normal luminosity during the plateau and a reasonable distance estimate, but appears to have very faint exponential decay phase. The reason is that the exponential decay is covered only by the ROTSE data, which are available only for this supernova. As a result, the coefficients $\beta_{n,i}$ at low $\tau$ are not directly connected to the rest of the data and have the peculiar values discussed in Section~\ref{sec:fits}. The situation could be easily rectified if the exponential decay in SN2006bp was observed in at least one photometric band with good coverage for other supernovae.

\subsection{Nickel masses}
\label{sec:mni}

\begin{figure*}
\includegraphics[height=6.5cm]{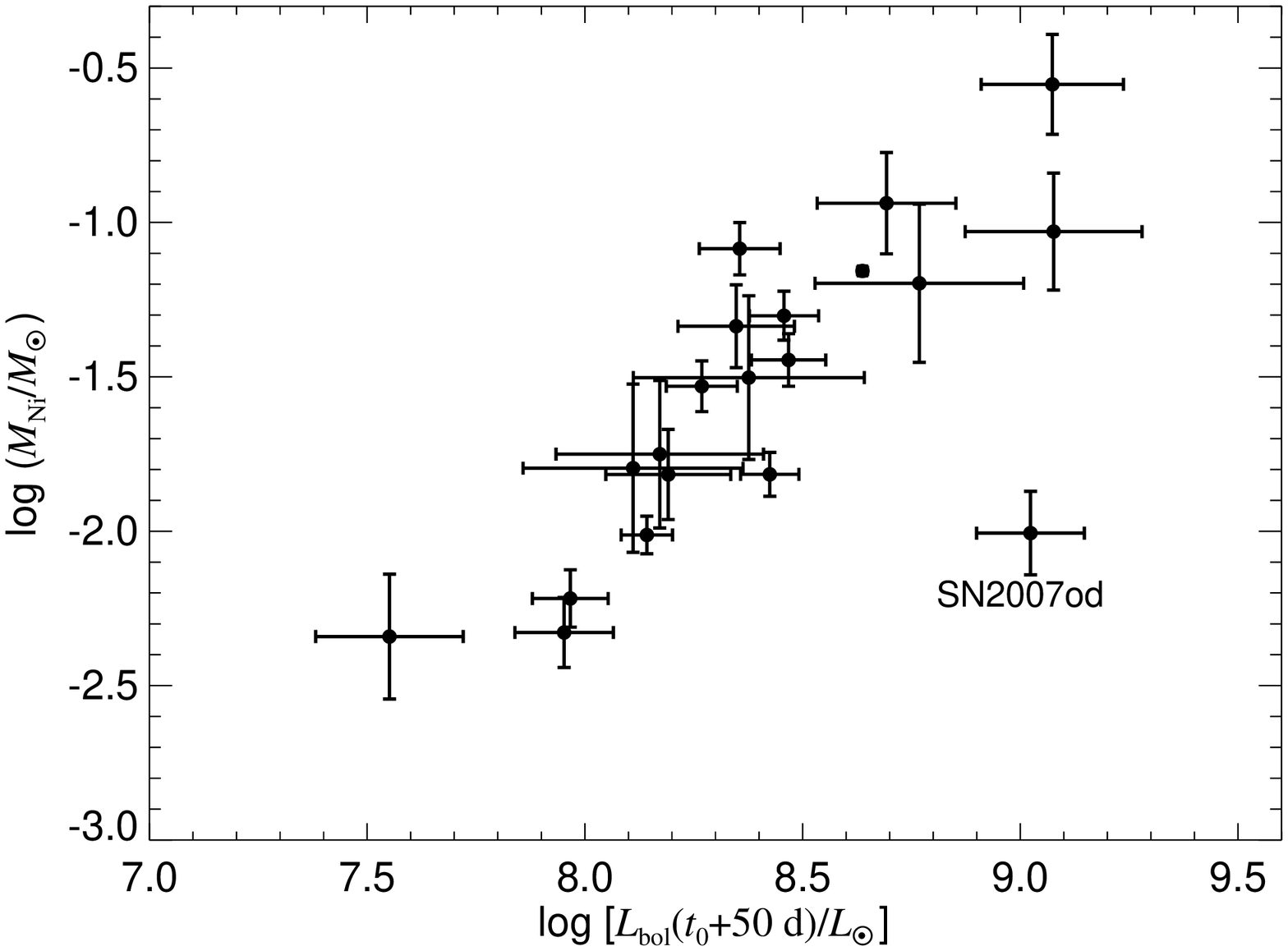}\hfill\includegraphics[height=6.5cm]{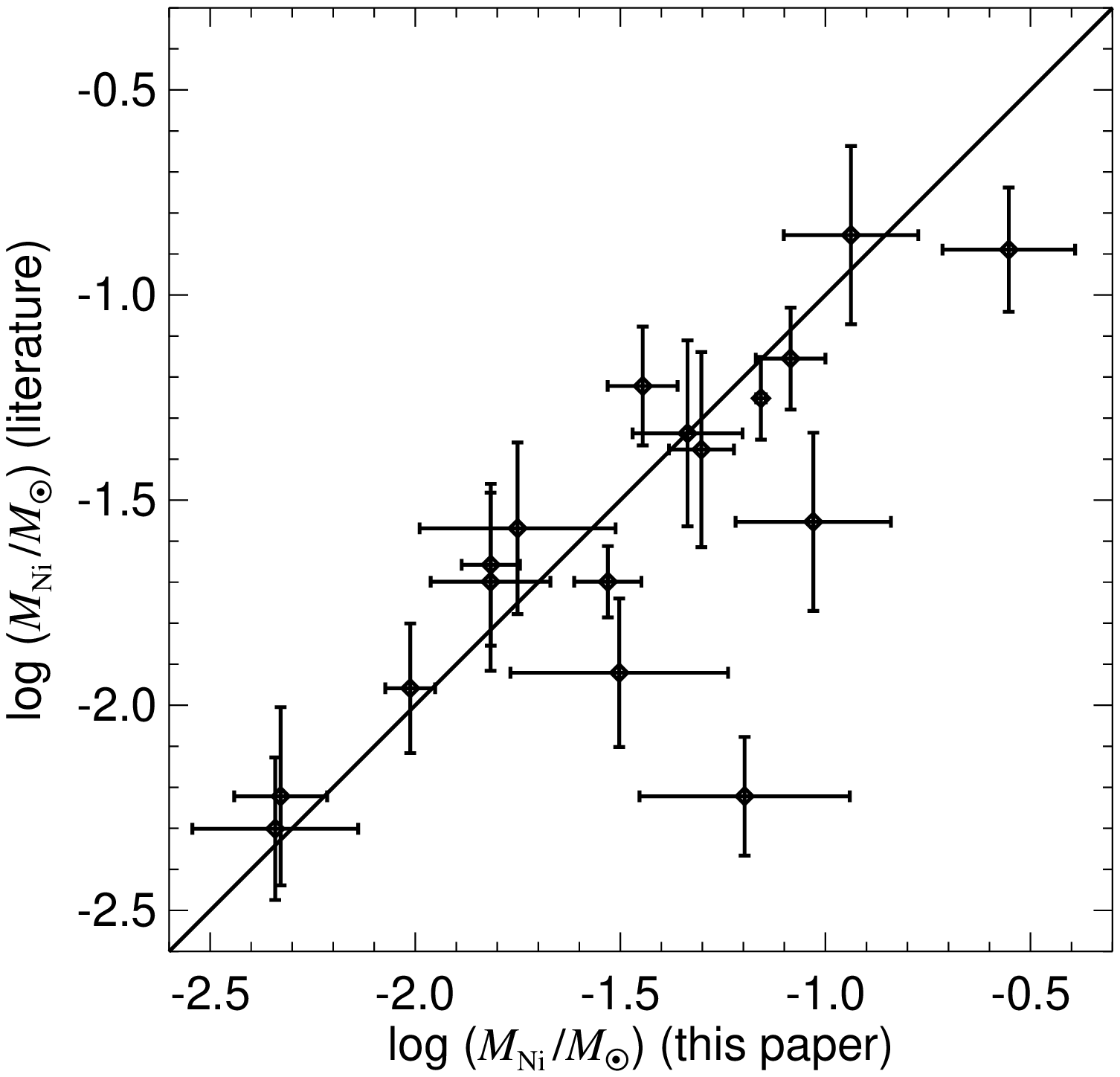}
\caption{{\em Left}: Nickel mass as a function of plateau $\lbol$ evaluated $50$\,days after $\texp$. Only supernovae with observations spanning before $50$\,d and after $200$\,d after $\texp$ are included. {\em Right}: Comparison of our $\mni$ with available results from the literature. We use $\mni$ determined by \citet{dallora14}, \citet{gandhi13}, \citet{hamuy03}, \citet{hendry06}, \citet{inserra12a,inserra13}, \citet{pozzo06}, \citet{sahu06}, \citet{spiro14}, \citet{tomasella13}, and \citet{vinko06}. Uncertainties on our results were rescaled to give $\mathcal{H}/\ndof = 1$. We do not show SN2006bp in either panel due to problems with the ROTSE temperature coefficients discussed in the text. }
\label{fig:mni}
\end{figure*}

\begin{deluxetable}{ccc}
\tabletypesize{\footnotesize}
 \tablewidth{0pc}
\tablecaption{Plateau luminosities and nickel masses}
\tablehead{ \colhead{Supernova}  & \colhead{$\log \left [\lbol(\texp+50\,{\rm d})/L_\sun \right]$} & \colhead{$\log (\mni/\msun)$}}
\startdata
SN2005cs   & $  7.952 \pm 0.045 $  &  $ -2.328 \pm 0.045 $ \\
SN2012A    & $  8.142 \pm 0.018 $  &  $ -2.012 \pm 0.018 $ \\
SN2004et   & $  8.468 \pm 0.027 $  &  $ -1.445 \pm 0.027 $ \\
SN2004A    & $  8.348 \pm 0.038 $  &  $ -1.336 \pm 0.039 $ \\
SN1999em   & $  8.457 \pm 0.023 $  &  $ -1.302 \pm 0.023 $ \\
SN2012aw   & $  8.638 \pm 0.003 $  &  $ -1.157 \pm 0.003 $ \\
SN2009bw   & $  8.425 \pm 0.020 $  &  $ -1.816 \pm 0.021 $ \\
SN2008in   & $  8.376 \pm 0.258 $  &  $ -1.503 \pm 0.258 $ \\
SN2007od   & $  9.023 \pm 0.039 $  &  $ -2.006 \pm 0.042 $ \\
SN2009js   & $  8.768 \pm 0.107 $  &  $ -1.197 \pm 0.117 $ \\
SN2004dj   & $  8.192 \pm 0.053 $  &  $ -1.816 \pm 0.054 $ \\
SN2002hh   & $  8.356 \pm 0.029 $  &  $ -1.085 \pm 0.027 $ \\
SN2008bk   & $  8.111 \pm 0.197 $  &  $ -1.796 \pm 0.200 $ \\
SN1992H    & $  9.074 \pm 0.056 $  &  $ -0.553 \pm 0.056 $ \\
SN2009N    & $  8.269 \pm 0.023 $  &  $ -1.531 \pm 0.023 $ \\
SN2001dc   & $  7.551 \pm 0.068 $  &  $ -2.341 \pm 0.078 $ \\
SN2006bp  \tablenotemark{a} & $  8.595 \pm 0.033 $  &  $ -2.822 \pm 0.069 $ \\
SN2009dd   & $  8.172 \pm 0.121 $  &  $ -1.750 \pm 0.120 $ \\
SN1995ad   & $  9.077 \pm 0.084 $  &  $ -1.030 \pm 0.078 $ \\
SN1996W    & $  8.693 \pm 0.049 $  &  $ -0.938 \pm 0.053 $ \\
SN1980K    & $  7.967 \pm 0.027 $  &  $ -2.218 \pm 0.029 $ \\

\enddata
\tablenotetext{a}{The absolute value of the nickel mass of SN2006bp is uncertain due to inadequately constrained temperature coefficients of the ROTSE bands.}
\label{tab:mni}
\end{deluxetable}

We can also use the bolometric light curves to calculate the ejected nickel mass $\mni$. We evaluate the luminosity at $200$ days after $\texp$ so that 
\beq
\mni = 4.494\times 10^{-43} \lbol(\texp+200\,{\rm days}) \msun,
\label{eq:mni}
\eeq
where the numerical coefficient is from \citet{hamuy03}. We show our $\mni$ as a function of $\lbol(\texp+50\,{\rm d})$ in the left panel of Figure~\ref{fig:mni} and we give the numerical values in Table~\ref{tab:mni}. We show only supernovae with observations before $50$\,days and after $200\,$days after $\texp$ to prevent extrapolation of the model. Uncertainties in both quantities are calculated using the full covariance matrix according to Equation~(\ref{eq:uncert_prop}) and we show the modified fit that gives $\mathcal{H}/\ndof = 1$. One potential caveat with Equation~(\ref{eq:mni}) is that many supernovae show faster exponential decay than what would be predicted by assuming full thermalization of the radioactive emission \citep[our Figs.~\ref{fig:lc}, \ref{fig:cor}, and Table~\ref{tab:sn_pars2} and also][]{anderson14a}. More realistic estimates are of $\mni$ are beyond the scope of this paper.

We recover the well-known correlation between plateau luminosity and nickel mass \citep[e.g.][]{hamuy03,spiro14}. The one exception is SN2007od, where the observations show a faint exponential decay phase. This has been explained by extinction due to dust formed in the supernova \citep{andrews10,inserra11} and we thus miss the infrared flux by assuming that the exponential decay colors of this supernova are the same as all other supernovae in our sample. It is interesting to note that the slope of the correlation in Figure~\ref{fig:mni} is almost exactly unity. The correlation is visible in supernovae with well-known distances and reddenings so it cannot just be a result of observational uncertainties in determining the luminosity. However, it would be more useful to study the ratio of the two quantities, which is insensitive to distance and reddenings errors.

In the right panel of Figure~\ref{fig:mni}, we show the comparison of our $\mni$ to previous results in the literature, where such numbers were easily available. There is an overal good agreement between our values and previous results except for several outliers. The outliers where we predict higher $\mni$ than what was obtained before are usually supernovae with a gap in coverage between the end of the plateau and late times of few hudred days after $\texp$. Note that these objects are not outliers with respect to the correlation with the plateau luminosity shown in the left panel of Figure~\ref{fig:mni}.

\subsection{Bolometric corrections}
\label{sec:bc}

A quantity of interest for observations of supernovae is the bolometric correction BC$_j$ with respect to band $j$. In our formalism,
\beq
{\rm BC}_j = -2.5\log \mathcal{F}\bol - \overline{M}_j + 2.5\Theta_j + \mathcal{C}\bol,
\label{eq:bc}
\eeq
where $\mathcal{C}\bol$ is the bolometric flux zero point. We choose
\beq
\mathcal{C}\bol = 2.5\log \left[\frac{L_{{\rm bol},0}}{4\pi (10\,{\rm pc})^2}\right] \approx -11.48,
\eeq
where $L_{{\rm bol},0} = 3.055 \times 10^{35}$\,\ergs\ based on the recommendation of the International Astronomical Union \citep{andersen99}\footnote{The reference is not available in the ADS and we thus use the value from Eric Mamajek's webpage \url{https://sites.google.com/site/mamajeksstarnotes/bc-scale} \citep{mamajek12,pecaut13}.} . This definition of $\mathcal{C}\bol$ does not require integrating synthetic photometry or theoretical spectra. \citet{bersten09} used $\mathcal{C}\bol \approx -11.64$, which they obtained by integrating the SED of Vega; the difference of $0.16$\,mag is an estimate of systematic uncertainty in BC$_j$ coming from slightly different procedures to obtain bolometric magnitude.

\begin{figure*}
\centering
\includegraphics[width=0.32\textwidth]{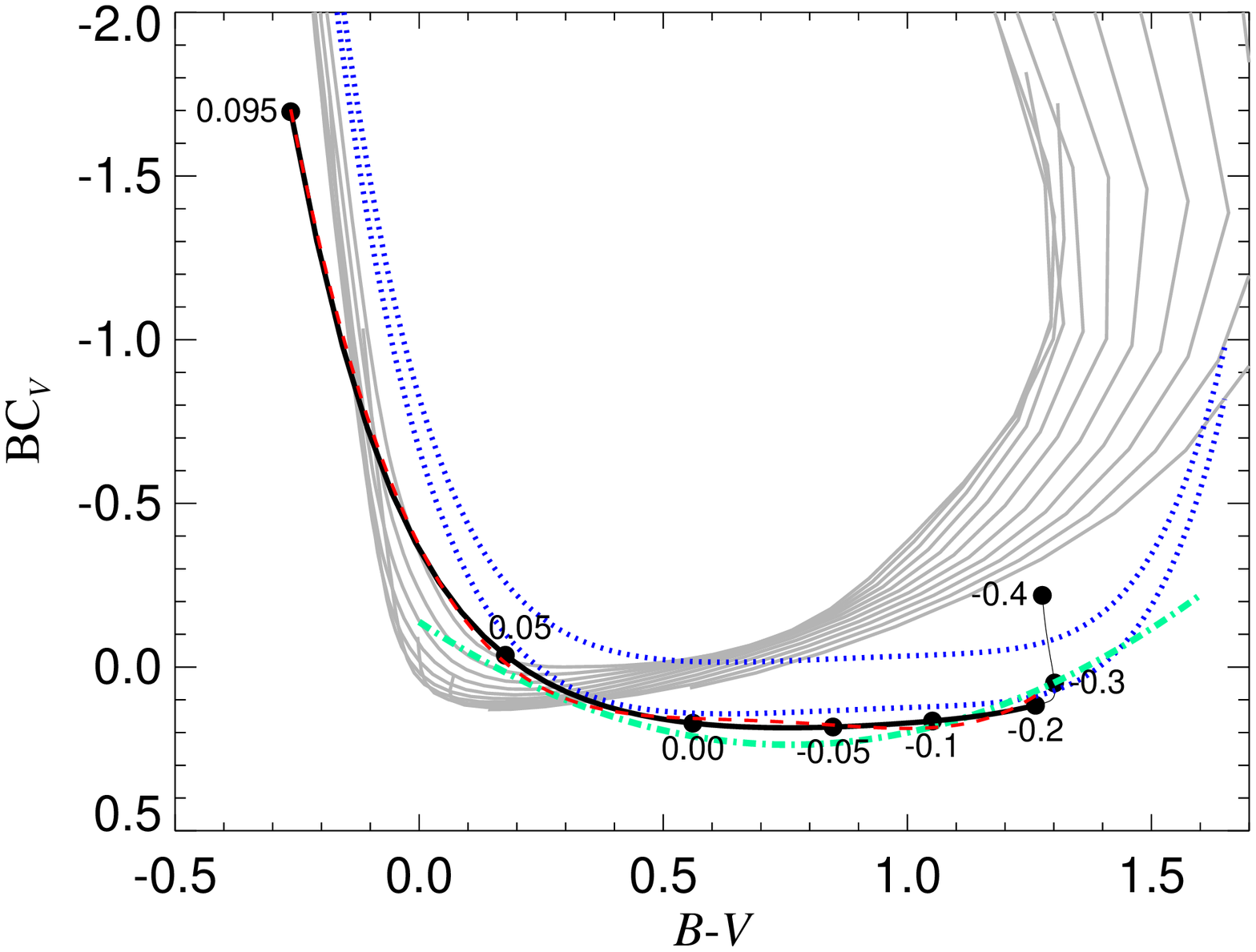}
\includegraphics[width=0.32\textwidth]{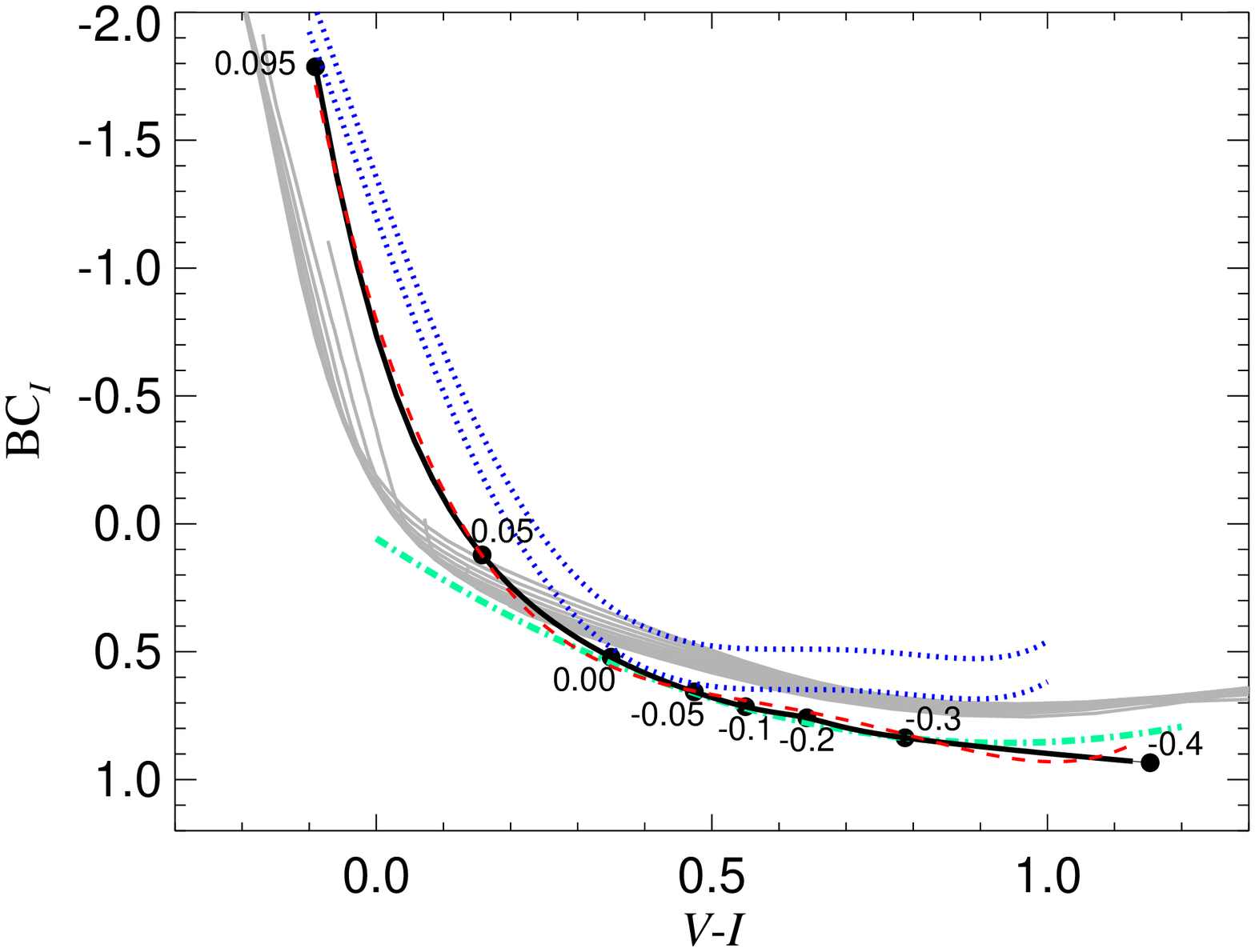}
\includegraphics[width=0.32\textwidth]{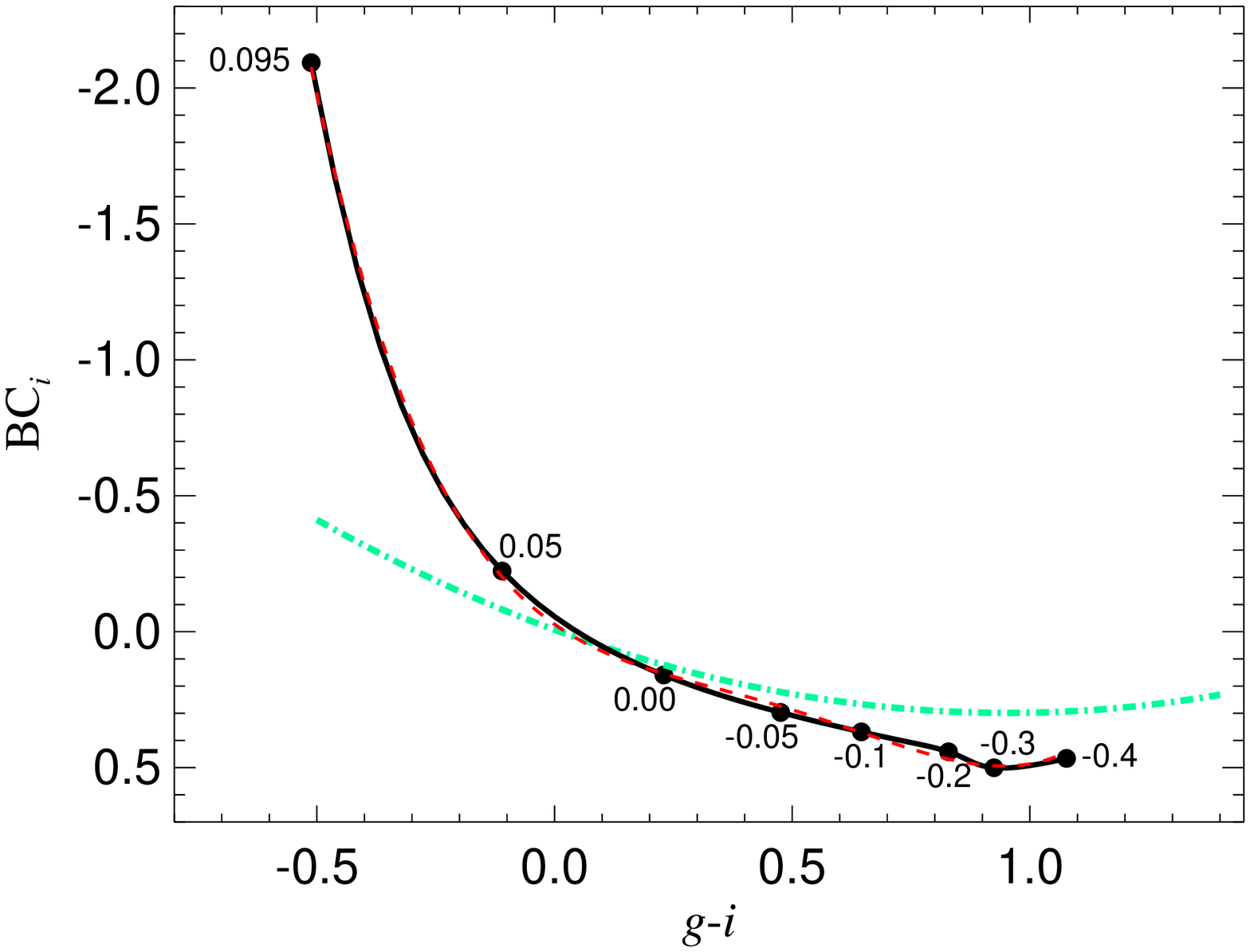}
\caption{Bolometric corrections to the $V$ band as a function of $B-V$ (left panel), to the $I$ band as a function of $V-I$ (middle), and to the $i$ band as a function of $g-i$ (right). This solid lines show calculations from our model and the range of colors used in fitting is indicated with thick solid line. Several values of $\tau$ are denoted by solid circles. The best-fit polynomials are shown with red dashed lines. The blue dotted lines show the bolometric corrections estimated for Type II-P supernovae by \citet{bersten09} with the original zero point (upper line) and our zero point (lower line). The green dash-dotted line shows the estimates of \citet{lyman14}, and the gray lines show bolometric corrections for normal stars with varying surface gravity and effective temperature \citep{bessell98}.}
\label{fig:bc}
\end{figure*}

In Figure~\ref{fig:bc}, we show bolometric corrections calculated from Equation~(\ref{eq:bc}) for the three color indices. We see that for blue colors, corresponding to high $\tau$ at early epochs, the bolometric corrections are quite significant. In Table~\ref{tab:bc_tran}, we give the mean color indices and bolometric corrections during the exponential decay phase corresponding to $\tau = -0.4$.

\begin{deluxetable}{ccc}
\tabletypesize{\footnotesize}
\tablecolumns{3}
\tablewidth{0pc}
\tablecaption{Mean colors and bolometric corrections during exponential decay}
\tablehead{ \colhead{}  & \colhead{Color} & \colhead{BC} }
\startdata 
  $U-B$ &   \phs $1.46$  &  $-1.49$ \\
  $B-V$ &  \phs $1.27$  &  $ -0.21$ \\
  $V-R$ &   \phs $0.80$  &\phs  $  0.59$ \\
  $R-I$ &  \phs $0.35$  & \phs $  0.93$ \\
  $I-J$ & \phs  $0.27$  &\phs  $  1.20$ \\
  $J-H$ &  $-0.02$  &   \phs $1.18$ \\
  $H-K$ & \phs  $0.34$  &\phs  $  1.52$ \\
  $g-r$ & \phs  $1.03$  & \phs $  0.42$ \\
  $r-i$ & \phs  $0.04$  & \phs $  0.47$ \\
  $i-z$ & \phs  $0.24$  & \phs $  0.70$
\enddata
\label{tab:bc_tran}
\end{deluxetable}

As expected, we find that our BC$_V$ is very different from normal stars (grey lines in Fig.~\ref{fig:bc}) since the SED is strongly modified by spectral lines and other physics. Our BC$_V$ is nearly identical to the one of \citet{bersten09} if we take into account the difference in zero point $\mathcal{C}\bol$. We also obtain a very good agreement with the bolometric corrections of \citet{lyman14}. For BC$_I$, we find reasonably good agreement with \citet{lyman14}, but the results of \citet{bersten09} start to deviate for $V-I \gtrsim 0.7$\,mag. For the Sloan bands, we show BC$_i$ as a function of $g-i$. Here, the agreement with \citet{lyman14} is noticeably worse, especially for $g-i \gtrsim 0.8$\,mag and $g-i \lesssim -0.3$\,mag. The reason might be that \citet{lyman14} obtained Sloan magnitudes by extracting fluxes at effective wavelengths of Sloan filters from SEDs constructed using Johnson bands. This does not fully take into account the presence and evolution of the spectral lines, which could affect color indices. For all cases shown in Figure~\ref{fig:bc}, \citet{lyman14} underpredict the bolometric correction right after explosion, when significant flux lies in near-UV. We capture some of this flux by including the \swift\ bands, although not all as discussed in Section~\ref{sec:sed} and Figure~\ref{fig:sed}. On the other hand, \citet{bersten09} make the UV correction based on theoretical models of supernova SEDs. The primary difference of our results with respect to the previous works of \citet{bersten09} and \citet{lyman14} is that we include the \swift\ near-UV bands and have a broader range of filters.

\begin{deluxetable*}{cccccccc}
\tabletypesize{\footnotesize}
\tablecolumns{8}
\tablewidth{0pc}
\tablecaption{Bolometric corrections}
\tablehead{ \colhead{Color}  & \colhead{Range} & \colhead{$c_0$}& \colhead{$c_1$}& \colhead{$c_2$}& \colhead{$c_3$}& \colhead{$c_4$} & \colhead{$\sigma$}}
\startdata
$U - B$ &  $(-1.27, 1.46)$ &  $ -0.3222 $ &  $ -0.3297 $ &  $  0.0383 $ &  $  0.2977 $ &  $ -0.4111 $ & $   0.162 $\\
$U - V$ &  $(-1.54, 2.74)$ &  $  0.0691 $ &  $  0.2461 $ &  $ -0.2797 $ &  $  0.1561 $ &  $ -0.0314 $ & $   0.021 $\\
$U - R$ &  $(-1.53, 3.53)$ &  $  0.1871 $ &  $  0.4333 $ &  $ -0.3031 $ &  $  0.1153 $ &  $ -0.0158 $ & $   0.024 $\\
$U - I$ &  $(-1.63, 3.87)$ &  $  0.1943 $ &  $  0.4891 $ &  $ -0.2744 $ &  $  0.0863 $ &  $ -0.0094 $ & $   0.031 $\\
$U - J$ &  $(-1.73, 4.17)$ &  $  0.2021 $ &  $  0.5331 $ &  $ -0.2434 $ &  $  0.0691 $ &  $ -0.0068 $ & $   0.033 $\\
$U - H$ &  $(-1.67, 4.09)$ &  $  0.2460 $ &  $  0.5701 $ &  $ -0.2461 $ &  $  0.0785 $ &  $ -0.0097 $ & $   0.020 $\\
$U - K$ &  $(-1.57, 4.38)$ &  $  0.2863 $ &  $  0.6475 $ &  $ -0.2749 $ &  $  0.0681 $ &  $ -0.0061 $ & $   0.035 $\\
$U - g$ &  $(-1.42, 2.34)$ &  $ -0.0373 $ &  $  0.0369 $ &  $ -0.3068 $ &  $  0.2238 $ &  $ -0.0575 $ & $   0.024 $\\
$U - r$ &  $(-1.64, 3.36)$ &  $  0.1052 $ &  $  0.3484 $ &  $ -0.2701 $ &  $  0.1104 $ &  $ -0.0159 $ & $   0.026 $\\
$U - i$ &  $(-1.93, 3.41)$ &  $ -0.0220 $ &  $  0.3069 $ &  $ -0.1836 $ &  $  0.0833 $ &  $ -0.0129 $ & $   0.024 $\\
$U - z$ &  $(-2.20, 3.64)$ &  $ -0.0837 $ &  $  0.3260 $ &  $ -0.1608 $ &  $  0.0572 $ &  $ -0.0063 $ & $   0.036 $\\
$uvw2 - U$ &  $(-1.37, 2.19)$ &  $  0.3854 $ &  $  0.1288 $ &  $ -0.1786 $ &  $  0.0098 $ &  $ -0.0046 $ & $   0.002 $\\
$uvm2 - U$ &  $(-2.12, 2.56)$ &  $  0.4056 $ &  $  0.0378 $ &  $ -0.0985 $ &  $  0.0023 $ &  $ -0.0014 $ & $   0.002 $\\
$uvw1 - U$ &  $(-1.11, 1.32)$ &  $  0.4058 $ &  $  0.0703 $ &  $ -0.3613 $ &  $  0.0144 $ &  $ -0.0198 $ & $   0.002 $\\
$u - U$ &  $(-0.64, 0.35)$ &  $  0.1829 $ &  $ -1.6143 $ &  $ -3.4435 $ &  $ -2.8074 $ &  $ -4.3188 $ & $   0.014 $\\
$U - b$ &  $(-1.23, 0.90)$ &  $ -0.2415 $ &  $ -0.3491 $ &  $ -0.2551 $ &  $  0.3189 $ &  $ -0.2645 $ & $   0.005 $\\
$U - v$ &  $(-1.58, 1.75)$ &  $  0.0601 $ &  $  0.2695 $ &  $ -0.2332 $ &  $  0.1355 $ &  $ -0.0414 $ & $   0.005 $\\
$B - V$ &  $(-0.26, 1.27)$ &  $ -0.3716 $ &  $  2.9669 $ &  $ -6.2797 $ &  $  5.8950 $ &  $ -2.0233 $ & $   0.023 $\\
$B - R$ &  $(-0.25, 2.06)$ &  $ -0.5660 $ &  $  3.1858 $ &  $ -3.8405 $ &  $  2.1115 $ &  $ -0.4220 $ & $   0.028 $\\
$B - I$ &  $(-0.35, 2.40)$ &  $ -0.4779 $ &  $  2.5717 $ &  $ -2.5096 $ &  $  1.1576 $ &  $ -0.1914 $ & $   0.030 $
\enddata
\tablecomments{Coefficients of a polynomial fit to the bolometric corrections, ${\rm BC}_j = \sum_{k=0}^4 c_k (m_i-m_j)^k$, where $\lambda_j > \lambda_i$. The fit is valid over a range of $m_i-m_j$ given in the second column. We also give the standard deviation about the fit $\sigma$, which only reflects how the fitting formula approximates BC$_j$ and not the true uncertainty in determining BC$_j$ from the observations. The Table is published in its entirety in the electron edition.}
\label{tab:bc}
\end{deluxetable*}

In order to facilitate supernova bolometric corrections for a wide range of filters, we fit a polynomial in color $m_i-m_j$ to BC$_j$ 
\beq
{\rm BC}_j = \sum_{k=0}^4 c_k (m_i-m_j)^k,
\label{eq:bcfit}
\eeq
where $m_j$ is always the redder band, $\lambda_i < \lambda_j$. We perform the fit only over a range of colors, $m_i-m_j$, where our model is valid. Specifically, we limit $\tau < 0.095$ and we remove the part of the curve, where BC$_j(m_i-m_j)$ is doubly-valued (e.g. $B-V>1.27$ for BC$_V(B-V)$ as indicated by the thick black line in Fig.~\ref{fig:bc}). If one of the filters is from \swift, we apply additional limit of $\tau > -0.06$ as discussed in Section~\ref{sec:fits}. The fit is performed on a uniform grid of $m_i-m_j$. The resulting coefficients of the fits, as well as the range of validity in $m_i-m_j$, are given in Table~\ref{tab:bc}. We also give the standard deviation of the residuals about the fit $\sigma$, although we emphasize that this measures only how well the fit approximates the theoretical curve and not the true uncertainty in determining BC$_j$ from the observations.

\begin{figure}
\plotone{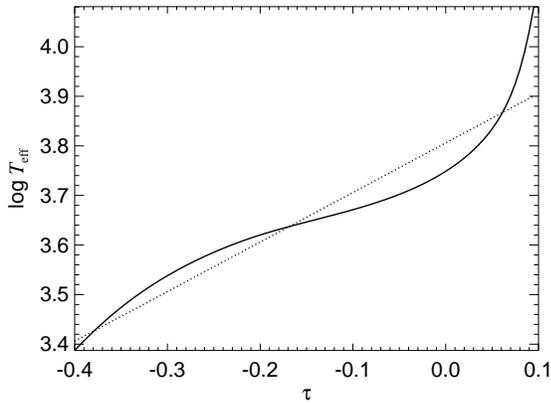}
\caption{Supernova effective temperature $T_{\rm eff}$ as a function of of our temperature parameter $\tau$ (solid line). Uncertainties in $T_{\rm eff}$ are $\lesssim 2\%$ from Equation~(\ref{eq:uncert_prop}). For comparison, we plot with dotted line a fit assuming linear dependence between $\tau$ and $\log T_{\rm eff}$.}
\label{fig:teff}
\end{figure}

Finally, we calculate the effective temperature $\teff$ from $\mathcal{F}\bol$ as
\beq
\teff = \left(\frac{R_0}{10\,{\rm pc}} \right)^{-\frac{1}{2}} \left(\frac{\mathcal{F}\bol}{\sigma_{\rm SB}}  \right)^{1/4},
\label{eq:teff}
\eeq
where $R_0=8.64\times 10^{9}$\,cm is the radius zero point corresponding to $\Pi =0$, and $\sigma_{\rm SB}$ is the Stefan--Boltzmann constant. In Figure~\ref{fig:teff} we show the relation between $\teff$ and our temperature parameter $\tau$. We see that the dependence between these two quantities is basically linear for $-0.4 \le \tau \lesssim 0.05$, as intended in our model (Sec.~\ref{sec:model}). The dependence steepens significantly for higher temperatures due to the strong near-UV flux.

\subsection{Dilution factors}
\label{sec:dilution}

\begin{figure*}
\plotone{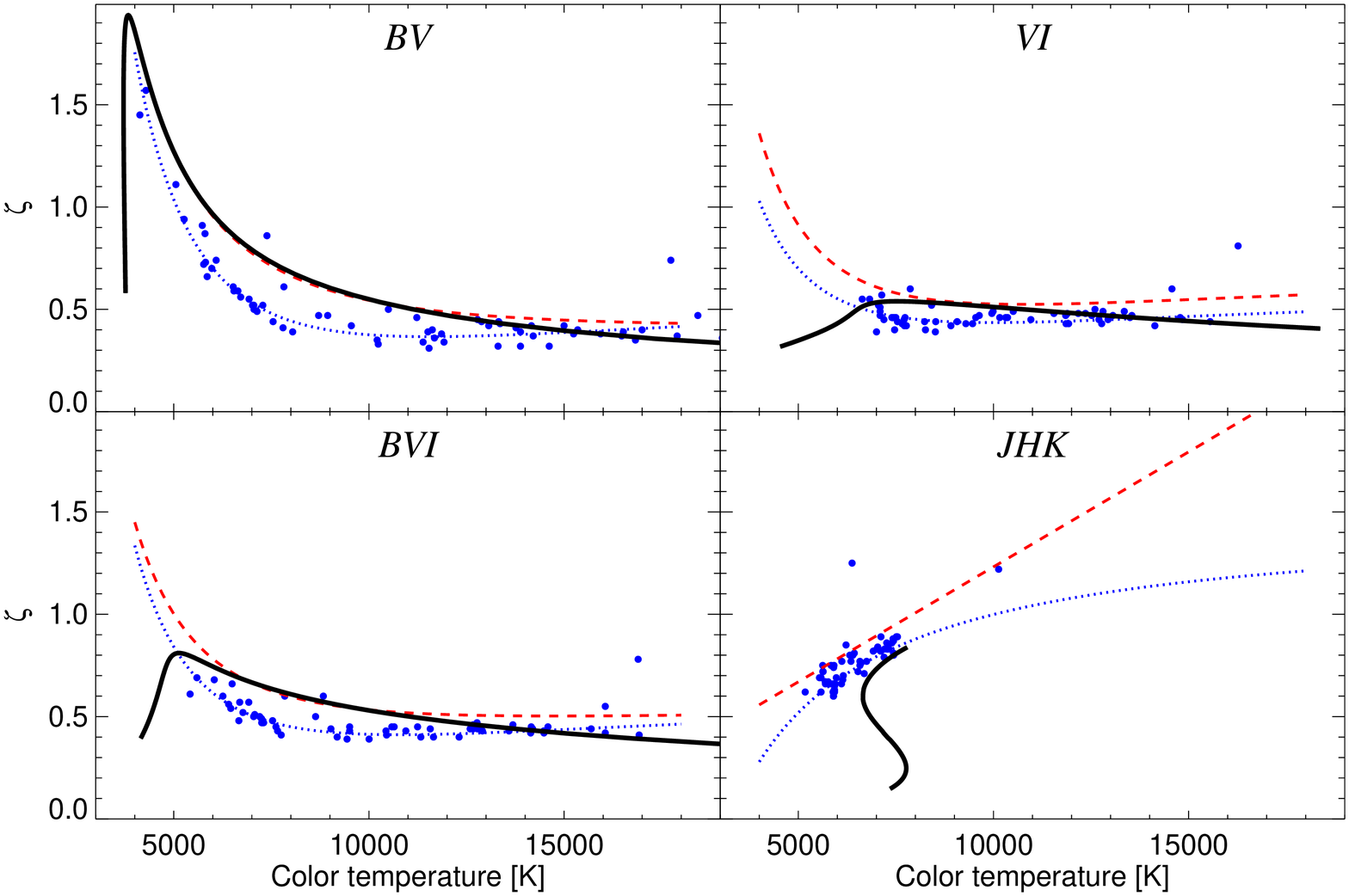}
\caption{Dilution factors $\zeta$ derived from our model (black lines) as a function of color temperature. We also show models of \citet[][blue points]{eastman96} along with their fit by \citet[][blue dotted lines]{hamuy01}, and similar fits to the models of \citet[][red dashed lines]{dessart05}.}
\label{fig:dilution}
\end{figure*}

Our model offers a unique way to empirically constrain the theoretical models of supernova atmospheres, which are a necessary component of the expanding photosphere method. In practice, theory supplies the ``dilution factors'' $\zeta$, which bring the flux produced by a black body with a color temperature $\tc$ derived from some combination of photometric bands to the true flux of the supernova. As a result, empirically determined values of $\zeta$ can be used as a check on the supernova atmosphere models.

We use the absolute magnitudes of our model as an input to the procedure of \citet{hamuy01}, which determines $\zeta$ and the color temperature for a given combination of filters. In Figure~\ref{fig:dilution}, we show our results for the filter combinations $BV$, $BVI$, $VI$, and $JHK$ along with the theoretical results of \citet{eastman96} and their fit by \citet{hamuy01}, and \citet{dessart05}. For the sake of completeness, we show the full range of temperatures covered by our model, $-0.4 \le \tau \lesssim 0.1$, which results in a break in $\zeta$ at low color temperatures due to the transition to the optically-thin exponential decay with its low color temperature (Fig.~\ref{fig:sed}). The low-temperature parts of the supernova light curves are not used for the expanding photosphere method and $\zeta$ in these areas is thus not of interest. We also restrict the comparison to the range of colors temperature actually covered by the models. We explicitly show the individual results of \citet{eastman96} with blue points. The ranges of color temperatures in \citet{dessart05} and \citet{dh08} are slightly greater.

For the $BV$ filters, our results agree extremely well with \citet{dessart05} for $4000 \le \tc \le 12000$\,K and begin to deviate slightly at higher temperatures. The models of \citet{eastman96} are systematically lower by a relatively large factor (an offset in $\zeta$ translates directly to an offset in the linear distance). For $BVI$, the range of $\tc$ with good agreement with \citet{dessart05} is smaller and \citet{eastman96} models again exhibit an offset. For $VI$, the agreement with \citet{dessart05} is reasonably good around $\tc \approx 9000$\,K, but we obtain decreasing $\zeta$ for both higher and lower temperatures. The \citet{dessart05} models with $\tc \approx 6000$\,K have $\zeta \approx 0.8$ and the time-dependent effects studied by \citet{dh08} lower $\zeta$ to about $0.6$, both of which are still higher than $\zeta \approx 0.4$ from our model. However, such low $\tc$ correspond to the optically-thin exponential decay in our model. This disagreement is probably due the spectral features in the $I$ bands, because $BV$ shows perfect agreement with the models, while the differences start to show up in $BVI$, where the $I$ band plays subdominant role for $\tc$ and $\zeta$ determination. Finally, for $JHK$ our results are systematically offset from \citet{dessart05}\footnote{Note that \citet{dessart05} fitting coefficients for $JHK$ filter combination apparently correspond to a polynomial in $T$ rather than $1/T$, unlike all the other cases. This can be seen by comparing their Figure~1 and Table~1.} and \citet{eastman96} at $\tc \approx 7000$\,K, and the dependence on $\tc$ is also completely different. This is probably due to the fact that $JHK$ exhibit similar flux in SEDs with very different blue and near-UV behaviors (Fig.~\ref{fig:sed}). This suggests that the range of $JHK$ color temperatures that can be used for distance determination with the expanding photosphere method is very small. Still, near-IR measurements can be useful in the standardized candle method, because the observed magnitude is more proportional to the radius of the supernova ($\Theta_i$ is small) and uncertainties in reddening play smaller role \citep[e.g.][]{maguire10a}.

We note that the calculation of $\zeta$ within our model did not require any input other than data in Table~\ref{tab:global} and the relation between magnitude and black-body color temperature of \citet{hamuy01}, which is used also in \citet{dessart05}. We would like to emphasize that the usual expanding photosphere method is limited by the subsets of filters, which have precalculated dilution factors. The model presented in this paper can be used to derive relative distances with any combination of filters from our Table~\ref{tab:global} and new photometric bands are easy to add.

\section{Discussions}
\label{sec:disc}

\subsection{Reddening law}
\label{sec:reddening}

In this paper, we have so far assumed the \citet{cardelli89} reddening law with $\rv=3.1$. In principle, we can fit for $\ri$ within our model similar to what we did for Cepheids in \citet{pejcha12}. However, we find that this procedure gives unreliable results, especially for bands with incomplete coverage of the full range of $\tau$ such \swift\ bands and Sloan $iz$. Instead, we vary only $\rv$ and assign $\ri$ to other bands based on the \citet{cardelli89} law. For all our data, the best-fit value is $\rv = 3.67 \pm 0.04$ and the fit improves by $\Delta \mathcal{H} \sim 1000$. We repeated the fits without including any \swift\ photometry and we find  $\rv = 2.90 \pm 0.03$, which is noticeably lower than what we obtained with \swift\ bands included. In either case, the improvement in the individual fits over the standard $\rv = 3.1$ is not dramatic indicating that the mean reddening law to our sample is compatible with the standard one. With more data, our model holds a good promise in determining the reddening law coefficients $\ri$ toward supernovae without spectroscopy or further assumptions. We emphasize that we assume a single mean reddening law towards all supernovae and that we do not differentiate between contributions from the Galaxy and the supernova host galaxy or the circumstellar medium.

\subsection{Possible extensions}
\label{sec:extensions}

Our model can be extended in other ways. Most importantly, K-corrections can be added to properly treat supernovae with non-negligible redshift. The model overpredicts the strength of the bump at the end of the plateau for the three bluest \swift\ bands. In this case, a higher-order expansion of the temperature coefficients (Eq.~[\ref{eq:theta}]) would be beneficial, but there is little data even for the best-observed supernova, SN2012aw. A somewhat worse fit is obtained for parts of the exponential decay phase, when we assume that the supernova colors are constant. However, there can be noticeable color evolution either immediately after the transition (SN2005cs) or gradually during longer time spans (SN2004et). The model predicts too bright $U$-band magnitude during the exponential decay phase in SN1999em, but this might be attributed to a single imprecise observations, because the agreement is much better in other objects (SN2012aw) with more data.  Clearly, a better model of the nebular phase spectrum would be appropriate, although we note that it might be difficult to constrain it with data since supernovae are usually already quite faint this late in their evolution.

We experimented with extending our coverage of photometric bands to Spitzer IRAC bands using photometry published by \citet{gandhi13} and \citet{kotak05}. Unfortunately, the small number of measurements and their timing did not allow us derive robust values of the global parameters. Nonetheless, these bands occupy the Rayleigh-Jeans tail of the SED, which typically provides only a small correction to the overall bolometric light curve. Similarly, the small number of measurements did not allow us to include the Sloan $u$ band.

Another possible avenue to explore are the relations between expansion velocities derived from different spectral lines. We tied our model to the Fe II line at $5169$\,\AA, but we collected from papers also velocities measured on H$\alpha$, H$\beta$, H$\gamma$, Si, Sc, N, He, and other elements. Our model can be modified to provide global transformations between velocities measured on these individual lines without requiring simultaneous observations of a single supernova and use of all of these velocities for distance determinations. We plan to address this issue in a future work.

\section{Conclusions}
\label{sec:conclusions}

We presented a model that disentangles the observed multi-band light curves and expansion velocities of Type II-P supernovae into radius and temperature changes (Eqs.~[\ref{eq:main}--\ref{eq:theta}]). We applied the model to a dataset of $\sim 230$ Fe~II expansion velocities and $\sim 6800$ photometric measurements in $21$ photometric bands spanning wavelengths between $0.19$ and $2.2\,\mu$m for 26 supernovae. We performed a detailed investigation to ensure that the radius and temperature functions have the desired meaning (Fig.~\ref{fig:platlin}) and that the global parameters of the model exhibit the expected trends with wavelength (Tab.~\ref{tab:global} and Fig.~\ref{fig:koef}). Our findings can be summarized as follows
\begin{itemize}
 \item Supernova light curves are well described by changes in radius and temperature as evidenced by the fits (Fig.~\ref{fig:lc}). The light curve shape during the optically-thick plateau phase is determined by the interplay of the increasing photospheric radius and the decreasing temperature (Fig.~\ref{fig:platlin}), in the sense that faster plateau declines correspond to slower photospheric radius increases. This can explain both flat plateaus with bump just before the transition phase (SN2005cs) and the steep magnitude decline observed in Type II-Linear supernovae (SN1980K, Fig.~\ref{fig:sn1980k}). We show that the temperature evolution is very similar in all supernovae (Fig.~\ref{fig:temp}) and that the rate of radius increase is related to the exponent of the expansion velocity decay $\omega_1$ (Eq.~[\ref{eq:slope}]). We argue that $\omega_1$ is related to the density structure of the ejecta. The differences between supernovae thus reflect the structure differences in their progenitors. This explains the correlation between plateau magnitude and slope recently discussed by \citet{anderson14a}.
 \item We determined parameters for $26$ supernovae (Tabs.~\ref{tab:sn_pars} and \ref{tab:sn_pars2}), including explosion times, plateau durations, transition widths, and reddenings. We studied the mutual dependence of individual supernova parameters (Tabs.~\ref{tab:sn_pars} and \ref{tab:sn_pars2}; Fig.~\ref{fig:cor}) and the correlation matrix of the fit (Fig.~\ref{fig:covar}). We found that $E(B-V)$ is little correlated with other parameters implying the robustness of our reddening estimates. We determined distances to $23$ host galaxies (Tab.~\ref{tab:galaxy}) and found good agreement with previous results from Type II-P supernovae and other methods in most supernovae (Fig.~\ref{fig:dist_ned}). Our relative distance estimates do not require dilution factors.
 \item We constructed SEDs of supernovae covering the wavelength range between $0.19$ to $2.2\,\mu$m (Fig.~\ref{fig:sed}). Within our model, a full SED range can still be constructed even for a supernova observed only in a subset of bands. The information on missing bands is reconstructed based on the full global fit and tailored to each individual supernova through the temperature coefficients $\alpha_0$ and $\alpha_1$ unique to each supernova. We integrate the SEDs to produce bolometric light curves (Figs.~\ref{fig:lbol_err} and \ref{fig:lbol}) with uncertainties including the full covariance matrix of the model. We calculate ejected nickel masses (Tab.~\ref{tab:mni}) and reproduce the known the correlation with plateau luminosity (Fig.~\ref{fig:mni}). We calculate bolometric corrections for all our filters and provide convenient fitting formula as a function of color indices (Tab.~\ref{tab:bc}). Apart from small offset in zero point, our bolometric corrections agree relatively well with previous results (Fig.~\ref{fig:bc}). We also provide mean supernova colors and bolometric corrections during the exponential decay phase (Tab.~\ref{tab:bc_tran}).
 \item In order to compare our results to theoretical models of supernovae spectra and fluxes, we construct empirical dilution factors from our model. We find that the agreement with \citet{dessart05} models is very good for $BV$ bands, but becomes worse when $I$ band is included either in $BVI$ or $VI$ subset (Fig.~\ref{fig:dilution}). The agreement is worse at low or high temperatures, implying that dilution factors should not be used outside of their range of validity. Although the dilution factors near maximum light agree for $JHK$ bands, the theoretical and empirical trends are quite different. The models of \citet{eastman96} give systematically smaller dilution factors.
 \item As a proof of principle, we attempted to obtain the reddening law from our model. We find that bands with enough data do not differ dramatically from the standard \citet{cardelli89} reddening law with $\rv \sim 3.1$, but bands with fewer data (\swift, Sloan $iz$ and potentially also $JHK$) produce systematically smaller values of $\ri$. With more data, our model can provide better constraints on the reddening law towards Type II-P.
 \item We make the fitting code publicly available\footnote{\url{http://www.astro.princeton.edu/$\sim$pejcha/iip/}} along with the published light curves and velocities of SN2004et and SN2009N \citep{maguire10b,pritchard14,takats14}.
\end{itemize}


\section*{Acknowledgements}
We are grateful to Chris Kochanek for discussions and detailed comments on the manuscript. We thank Adam Burrows, Luc Dessart, and Dovi Poznanski for discussions and comments. We thank Todd Thompson for comments to the early version of the manuscript. We are grateful to Joe Anderson, John Beacom, Melina Bersten, Subo Dong, and the anonymous referee for comments that helped to improve the paper. We thank Andrea Pastorello, Kate Maguire, Katalin Tak\'ats, Stefano Valenti, Mario Hamuy, and Rupak Roy for providing data for various supernovae. We acknowledge The Weizmann interactive supernova data repository\footnote{\url{http://wiserep.weizmann.ac.il}}. Support for this work was provided by NASA through Hubble Fellowship grant HST-HF-51327.01-A awarded by the Space Telescope Science Institute, which is operated by the Association of Universities for Research in Astronomy, Inc., for NASA, under contract NAS 5-26555. This research has made use of the NASA/IPAC Extragalactic Database (NED), which is operated by the Jet Propulsion Laboratory, California Institute of Technology, under contract with the National Aeronautics and Space Administration.

\appendix
\section{Implementation of priors}

Here, we describe the implementation of Equation~(\ref{eq:constraint}) in {\tt cmpfit}. The fitting code {\tt cmpfit} does not provide an interface to implement non-trivial priors on fitted parameters. To obtain the fit, the user supplies a list of deviations of the data $y_i$ from the model $f_i$ weighted by the measurement uncertainty $\sigma_i$, $\delta y_i = (y_i-f_i)/\sigma_i$, where $i=1\ldots N_{\rm data}$ and $N_{\rm data}$ is the number of datapoints. To apply prior constraints, we append $\delta y_i$ with a vector $\delta p_j$, where $j=1\ldots N_{\rm prior}$ and $N_{\rm prior}$ is the number of priors. In the language of Equation~(\ref{eq:constraint}), $S=|\mathbf{\delta p}|^2$. A Gaussian prior on parameter $a_k$ with a mean value of $\bar{a}_k$ and width $\sigma_{a_k}$ is obtained by setting
\beq
\delta p_j = \frac{a_k-\bar{a}_k}{\sigma_{a_k}}.
\label{eq:app_diag}
\eeq
If parameter $a_k$ is unconstrained by the data applied on the model $f_i$, Equation~(\ref{eq:app_diag}) guarantees that after the fit $a_k$ will be equal to $\bar{a}_k$ with uncertainty $\sigma_{a_k}$.

Equation~(\ref{eq:app_diag}) needs to be modified to allow for ``non-diagonal'' priors, which prescribe correlations between individual parameters. Assume that the prior probability distribution of the parameter vector $\mathbf{a}$ is a Gaussian centered at $\mathbf{\bar{a}}$ with a covariance matrix $\mathbf{C} = \mathbf{PDP}^{-1}$, where $\mathbf{D}$ is a diagonal matrix of eigenvalues of $\mathbf{C}$ and $\mathbf{P}$ is the matrix of the corresponding eigenvectors. The deviations are thus
\beq
\mathbf{\delta p} = (\mathbf{a}-\mathbf{\bar{a}})\mathbf{PD}^{-1/2}.
\label{eq:app_nondiag}
\eeq
This form guarantees that if parameters $\mathbf{a}$ are not constrained by the data, their values after the fit will be $\mathbf{\bar{a}}$ with the covariance matrix $\mathbf{C}$. The diagonal elements of $\mathbf{C}$ are squares of the parameter uncertainties and Equation~(\ref{eq:app_nondiag}) reduces to Equation~(\ref{eq:app_diag}) for diagonal $\mathbf{C}$ with diagonal elements $\sigma_{a_k}^2$.

We use the priors of the form of Equation~(\ref{eq:app_diag}) to constrain values of $\mathscr{R}_i$ to within $10\%$ of the \citet{cardelli89}, when being fitted (Sec.~\ref{sec:reddening}). We use priors of the form of Equation~(\ref{eq:app_nondiag}) to constrain parameters of individual supernovae, because they can constrain possible correlations between individual parameters and provide a better fit. In a sense, $\mathbf{C}$ in Equation~(\ref{eq:app_nondiag}) is a mathematical representation of Figure~\ref{fig:cor}.

We apply priors for a subset of individual supernova parameters $\mathbf{a} = \{\tp, \tw, \omega_0, \omega_1, \omega_2, \alpha_0, \alpha_1, \gamma_0, \gamma_1\}$, where the mean $\mathbf{\bar{a}}$ and the covariance matrix $\mathbf{C}$ is obtained from supernovae with more than $5$ velocity measurements. Naturally, these priors will only be important for supernovae with data insufficient in some aspects, for example, small number of expansion velocities or no velocity measurements at all. In other cases, the priors will be completely overwhelmed by the data.

\end{document}